\documentclass[letterpaper,twocolumn,10pt]{article}
\usepackage{usenix}

\usepackage[normalem]{ulem}

\newcommand{\baidu}[0]{\texttt{Baidu}}
\newcommand{\metadb}[0]{\textit{MetaDB}}

\newcommand{\tidb}[0]{\textit{TiDB}}
\newcommand{\bos}[0]{\texttt{BOS}}
\newcommand{\cfs}[0]{\texttt{CFS}}
\newcommand{\mantle}[0]{\texttt{Mantle}}

\newcommand{\create}[0]{\texttt{create}}
\newcommand{\mkdir}[0]{\texttt{mkdir}}
\newcommand{\rmdir}[0]{\texttt{rmdir}}
\newcommand{\rename}[0]{\texttt{rename}}
\newcommand{\scan}[0]{\texttt{listdir}}

\newcommand{\dirstat}[0]{\texttt{dirstat}}
\newcommand{\delete}[0]{\texttt{delete}}

\usepackage{tikz}
\usepackage{amsmath}

\usepackage{filecontents}
\usepackage{subcaption}
\usepackage{caption}
\usepackage{graphicx}
\usepackage{enumitem} 
\usepackage{multirow}
\usepackage{booktabs}
\usepackage{makecell}

\usepackage{hhline} 
\usepackage[affil-it]{authblk}
\makeatletter
\renewcommand\AB@affilsepx{\quad\quad \protect\Affilfont}
\def\thanks#1{\protected@xdef\@thanks{\@thanks
        \protect\footnotetext{#1}}}
\makeatother

\author[1,*]{Zebin Ren}\thanks{*Equal contribution.}
\author[2,*,$\dagger$]{Biao Cao}
\author[
  3,
  $\dagger$\thanks{$\dagger$ Corresponding author.},
  $\ddagger$\thanks{$\ddagger$ Work done while the author was at Tsinghua University.}
]{Kang Chen}
\author[2]{Zhihui Yin}
\author[2]{Qiushi Chen}
\author[2]{Jielong Jian}
\author[2]{Jiwei Xiong}
\author[2]{Xinxing Wang}
\author[2]{Yan Xing}
\author[2]{Liguo Duan}
\author[2]{Ran Zheng}
\author[2]{Xianjun Meng}
\author[2]{Zhenyu Hou}

\affil[1]{Vrije Universiteit Amsterdam}
\affil[2]{Baidu (China) Co., Ltd.}
\affil[3]{Peking University}

\usepackage{listings}
\usepackage{xcolor}
\usepackage{caption}
\usepackage{tcolorbox}
\usepackage{pifont}

\usepackage{titlesec}

\titleformat{\subsubsection}[runin]
  {\bfseries}
  {\thesubsubsection}
  {0.5em}
  {}
  [.]

\definecolor{keywordcolor}{rgb}{0, 0, 0.5} 
\definecolor{commentcolor}{rgb}{0, 0.5, 0} 

\begin{document}

\date{}

\title{\Large \bf MetaDB: Putting File-System Structure Back into Distributed Databases}

\maketitle

\pagestyle{empty}

\begin{abstract}
Storage systems in production increasingly rely on distributed databases for metadata management, yet our experience at \baidu{} shows that treating a general-purpose database as a black box leaves substantial performance and efficiency untapped. File-system metadata exhibits directory-tree locality, differentiated consistency requirements, and clustered deletions that general-purpose databases largely ignore.
We thus propose \metadb{}, a metadata-aware distributed transactional database that serves as the shared metadata layer for \baidu{}’s large-scale storage systems. \metadb{} addresses three key mismatches between general-purpose databases and metadata workloads: Data Model Mismatch, Transaction Model Mismatch, and Storage Implementation Mismatch. To bridge these gaps, \metadb{} combines \ding{172} tree-topology-induced sharding and execution, \ding{173} semantic-aware off-critical-path processing, and \ding{174} a lifecycle-aware storage engine with in-memory MVCC and locality-aware compaction.

\metadb{} achieves 2.2$\times$--4.1$\times$ the throughput of our previous general-purpose database and halves the number of metadata machines.
Deployed in production at \baidu{} for over five years, \metadb{} unifies metadata management across the company’s storage fleet and scales to over 1{,}000 metadata nodes, tens of trillions of metadata records, and exabyte-scale storage in a single cluster.
\end{abstract}

\section{Introduction}
\label{sec:introduction}

Distributed storage systems are fundamental to modern cloud infrastructure, but efficiently managing massive metadata remains challenging. Many systems therefore adopt distributed relational or key-value databases for file-system metadata management (e.g., Tectonic~\cite{fast21-tectonic}, Colossus~\cite{colossus}, Mantle~\cite{sosp25-mantle}, ACOS~\cite{fast26-acos}). This decoupled architecture reduces development complexity by delegating state management, elasticity, and fault tolerance to database systems, while still providing high scalability and reliability.

At \baidu{}, we also use this paradigm to manage metadata for our exabyte-scale cloud storage infrastructure, including our cloud file system~(\cfs{}) and object storage services~(\bos{} and \mantle{})~(\S\ref{sec:mds-baidu}). We therefore built and deployed a unified general-purpose distributed database as the common metadata layer across all our storage systems.

However, our production experience shows that treating the database as a ``black box'' for metadata workloads leaves substantial performance untapped and causes severe bottlenecks. Because general-purpose databases are designed for generic relational or key-value data, they overlook the unique structural, semantic, and dynamic properties of file-system metadata. We identify three mismatches between general-purpose databases and metadata workloads.

First, the flat \textbf{data model} of traditional databases conflicts with the hierarchical tree structure of file-system metadata (\textit{Data Model Mismatch}). General-purpose databases typically rely on range or hash sharding, often splitting entries of the same parent directory across shards and thereby turning high-frequency single-directory updates into expensive distributed transactions. Even when locality is preserved within a shard, the database lacks directory-tree semantics, so a single logical operation (e.g., file creation) can still incur multiple network round-trips for the required transactional steps.

Second, the \textbf{generic transaction model} imposes a rigid, one-size-fits-all atomicity model for all records in a transaction. However, this model fails to accommodate the highly differentiated consistency requirements of metadata updates (Transaction Model Mismatch). Single metadata operations often trigger updates to parent-directory attributes and derived auxiliary metadata. Enforcing strict ACID guarantees across all these fields in a single transaction creates severe contention on shared parent-directory records. It also forces unnecessary multi-shard coordination for auxiliary metadata, even though such metadata can safely tolerate relaxed timeliness.

Third, the underlying \textbf{storage implementation} mismatches metadata access patterns (\textit{Storage Implementation Mismatch}). Traditional Multi-Version Concurrency Control~(MVCC) persists all tombstones to storage to support long-running transactions, whereas metadata transactions are short-lived ($<10$\,ms). Under cloud-scale consecutive bulk deletions, these persisted tombstones rapidly accumulate in the underlying Log-Structured Merge-tree~(LSM-tree) and degrade range scans such as \scan{}. Such deletions also exhibit strong spatial locality (e.g., tens of megabytes within a 4\,GB shard), but conventional compaction does not exploit this locality and thus cannot purge tombstones in time.

To bridge these gaps, we propose \metadb{}, a high-performance, metadata-aware distributed database at \baidu{}. Rather than treating the database as a generic black box, \metadb{} co-designs the database engine around metadata-specific structures, semantics, and dynamics through the following three techniques.

\textbf{\ding{172} Tree-Topology-Induced Sharding and Execution.} To resolve the data model mismatch, \metadb{} aligns database shard boundaries with the directory tree via policy/size-based sharding, keeping single-directory operations within a single shard. We further push down transactional logic directly to the storage nodes, collapsing multi-round network coordination into a single remote execution.

\textbf{\ding{173} Semantic-Aware Off-Critical-Path Processing.} To resolve the transaction model mismatch, \metadb{} extracts aggregatable hotspot directory attributes and auxiliary metadata from the critical path. For hotspot parent-directory attributes, synchronous in-place updates are removed from the write critical path and replaced with asynchronous aggregation through an in-memory queue plus read-path reconciliation, reducing lock contention while preserving the latest visible state. We employ an asynchronous off-critical-path update mechanism: auxiliary metadata (quotas and indexes) are updated out-of-band without strict timeliness requirements, thereby avoiding multi-shard transactions. 

\textbf{\ding{174} Lifecycle-Aware Storage Engine.} To resolve the storage implementation mismatch, \metadb{} optimizes the storage engine specifically for metadata lifecycles. By introducing an in-memory MVCC layer, we prevent short-lived MVCC tombstones from being flushed to disk. Combined with locality-aware range compaction, \metadb{} promptly purges localized tombstones, sustaining low latency for \scan{}.

By resolving the mismatches between databases and metadata workloads, \metadb{} delivers 2.2$\times$--4.1$\times$ the throughput of our previous general-purpose database. \metadb{} has been deployed in \baidu{}'s production environment for over five years, unifying metadata management across our entire storage fleet. \metadb{} reduces the number of cluster machines required for metadata management by approximately 50\%. Today, a single \metadb{} cluster scales to over 1{,}000 metadata machines, managing tens of trillions of records and supporting exabyte-scale storage.

This paper makes the following key contributions:
\begin{itemize}
    \setlength\itemsep{0pt}      
    \setlength\parskip{0pt}  
    \item We identify the root cause of metadata inefficiency in general-purpose distributed databases as the loss of structural information along three dimensions: tree topology, transaction semantics, and operation dynamics.
    \item We propose and implement three corresponding designs in \metadb{} to restore the tree structure in the database.
    \item We evaluate \metadb{} on massive-scale production metadata workloads, demonstrating substantial performance improvements and resource savings over general-purpose databases.
\end{itemize}

\section{Background and Motivation}
\label{sec:background}

\subsection{Metadata Management at \baidu{}}
\label{sec:mds-baidu}

In large-scale distributed storage, metadata encompasses the structural and attribute information governing data management. 
This specifically includes the \textit{flat and hierarchical namespaces}~(e.g., flat object keys and directory trees), \textit{system attributes}~(e.g., access control, modification timestamps, and user quotas), and auxiliary \textit{secondary indexes}~\footnote{Secondary indexes are utilized to accelerate queries on specific attributes, such as filtering objects by creation time for garbage collection.}.

The production environment at \baidu{} deploys two primary categories of storage infrastructure underpinned by distinct metadata management strategies:

\noindent
\textbf{Cloud Object Storage:} To accommodate diverse application demands, we provide two variants of namespace services. 
Object storage~(\bos{}) offers a \emph{flat namespace} ideal for key-value semantics. 
Hierarchical object storage~(\mantle{}~\cite{sosp25-mantle}) supports HDFS-like directory semantics. 
Architecturally, such a hierarchical namespace adopts a fully disaggregated design where clients are oblivious to data placement and cannot cache metadata locally. Thus, every metadata operation translates to a direct request to the backend metadata service, demanding high throughput and low latency.

\noindent
\textbf{Distributed File Storage:} Our distributed file storage~(\cfs{}~\cite{eurosys23-cfs}) provides a fully POSIX-compliant hierarchical namespace optimized for shared file access.
Unlike the hierarchical namespace for cloud object storage, this namespace for distributed file storage utilizes a coupled architecture where clients can cache directory metadata and physical layout information to accelerate local path resolution.

Despite these architectural differences in client-side caching and data layout, both \mantle{} and \cfs{} inherently require a robust and scalable storage layer to persist their hierarchical directory trees. Unless otherwise specified, we use \bos{}, \mantle{}, and \cfs{} in the remainder of this paper to refer specifically to their underlying \textit{namespace subsystems}.

\subsection{Metadata Databases Prior to \metadb{}}
\label{sec:prev-tafdb}

The distributed databases powering \baidu{}'s metadata services underwent two generations of architectural evolution before the inception of \metadb{}.

\noindent
\textbf{Generation 1: Ad-hoc Design with Multiple Databases.}
Initially, \bos{} and \cfs{} were served by separate metadata systems. For \bos{}, metadata was partitioned across $N$ MySQL shards using static modulo hashing for simplicity. This design had three major limitations:
(1) \textit{Inflexible Scalability}: static hashing forced an expensive ``double-the-cluster''~($N \rightarrow 2N$) upgrade path to avoid massive data migration;
(2) \textit{Lack of Atomicity}: multi-node operations, such as cross-shard \rename{}, lacked transactional guarantees and thus risked inconsistency; and
(3) \textit{High Management Overhead}: fragmented, system-specific metadata silos imposed high operational and maintenance costs.
Meanwhile, \cfs{} used static subtree partitioning, which was highly vulnerable to workload hotspots.

\noindent
\textbf{Generation 2: Unified Design with a General-Purpose Database.}
A dedicated metadata service (MDS) such as CephFS~\cite{osdi06-ceph} targets hierarchical namespaces and does not fit \bos{}'s flat namespace, whereas a distributed transactional database can serve both.
We therefore built a unified general-purpose distributed database, architecturally similar to TiDB~\cite{tidb} and CockroachDB~\cite{cockroachdb}, based on the open-source Doris project~\cite{doris}. It used RocksDB as the local storage engine beneath a distributed SQL proxy layer, and provided dynamic sharding, multi-Raft replication for high availability, and MVCC with snapshot isolation for distributed transactions.
While the system was functionally complete, production deployment exposed a key lesson: \textbf{treating the underlying database as a ``black box'' creates severe performance bottlenecks for metadata workloads.} 
Thus, we redesigned the database architecture around directory-tree semantics and built \metadb{}. Its design is motivated by three fundamental architectural mismatches we observed in Generation~2.

\subsection{Three Architectural Mismatches}
\label{subsec:motivation}

\subsubsection{Data Model Mismatch: Flat Sharding vs. Tree Topology}

\begin{figure}

\begin{minipage}{0.23\textwidth}
    \centering
    \includegraphics[width=\linewidth]{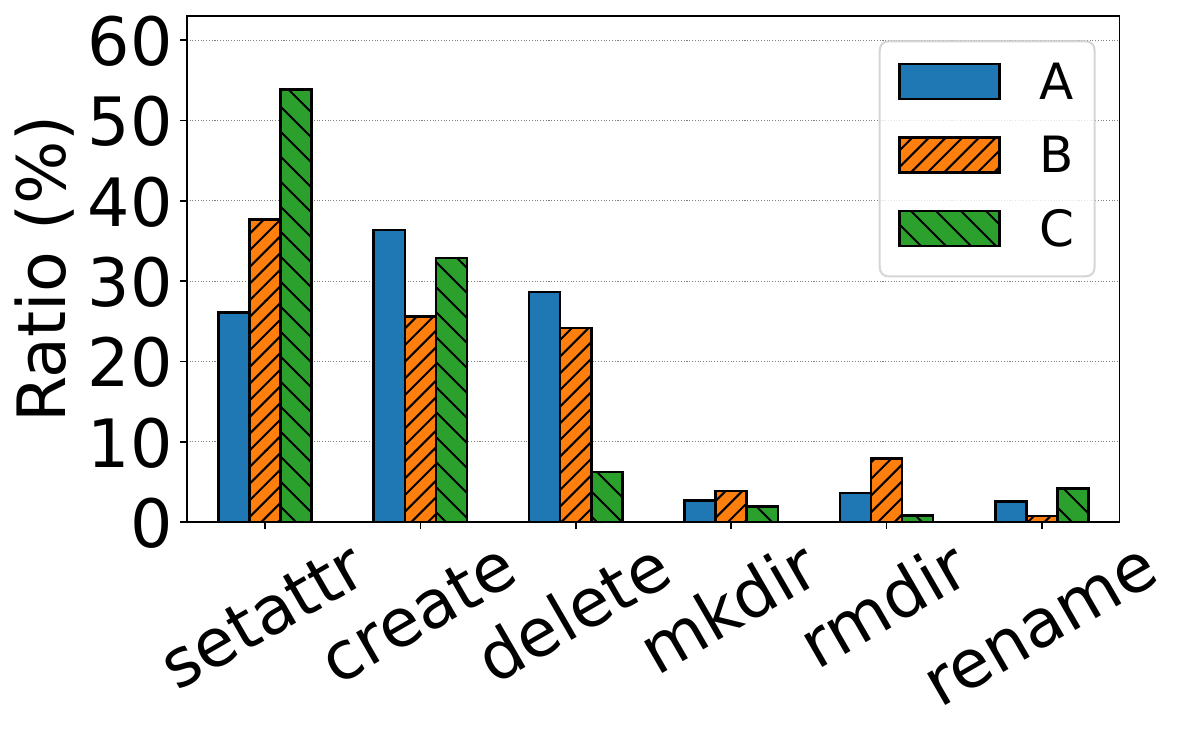}
    \caption{Op distribution.}
    \label{fig:workload-ops}
\end{minipage}
\hfill
\begin{minipage}{0.23\textwidth}
    \centering
    \includegraphics[width=\linewidth]{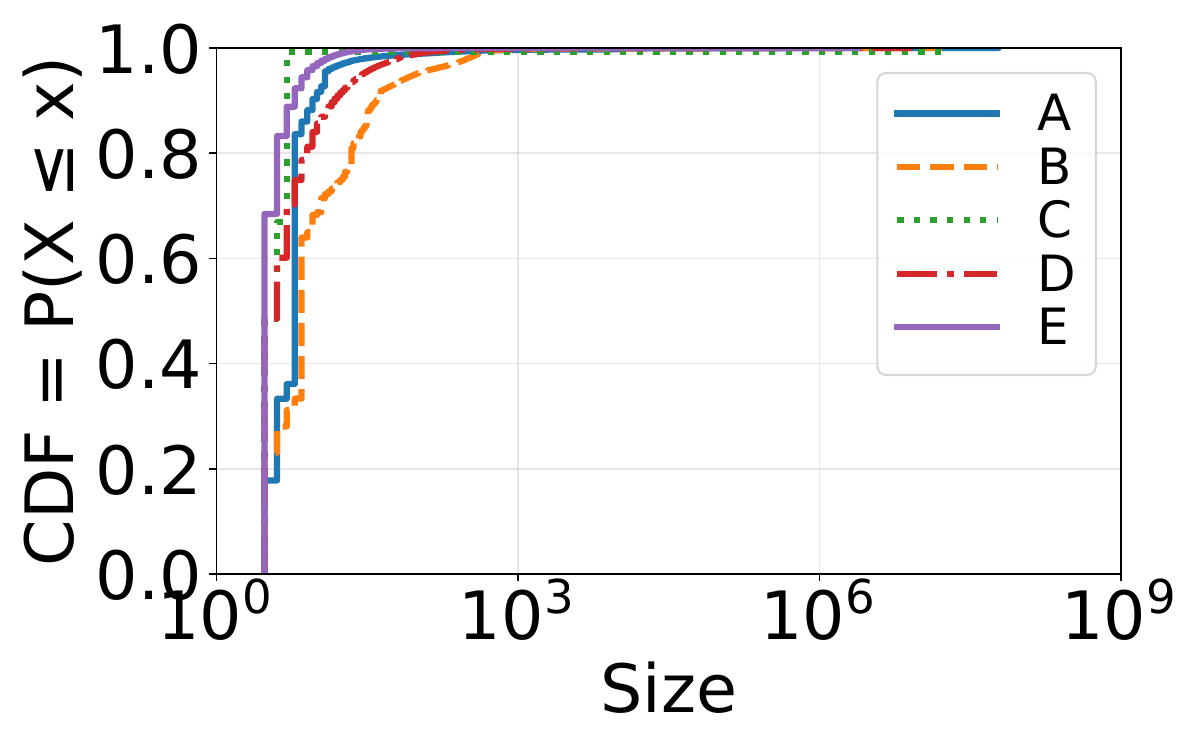}
    \caption{Directory sizes.}
    \label{fig:workload-dir-size}
\end{minipage}

\end{figure}
General-purpose databases view data as flat, independent records in tables, partitioning them by size or hash boundaries. This ignores the inherent topology of file systems.

\noindent
\textbf{Spatial locality is destroyed by flat sharding.} 
\label{sec:moti-op-dsitr}
\autoref{fig:workload-ops} breaks down directory-tree-modifying operations across three representative \cfs{} clusters, denoted by A, B, and C. Over 97\% of all operations (e.g., \mkdir{}, \create{}, \delete{}) are single-directory operations; cross-directory \rename{} accounts for less than 3\%. Logically, a single-directory operation should only access localized data. However, generic size-based sharding arbitrarily fractures large directories across multiple shards, forcibly escalating these high-frequency, localized updates into expensive distributed transactions.

\noindent
\textbf{Single-node containment is feasible.} \autoref{fig:workload-dir-size} shows the directory size distribution of five users. The majority of directories are bounded in size, with even the largest containing fewer than one billion entries. This suggests that aligning database shard boundaries with directory boundaries is viable and keeps single-directory workloads within a single shard.

\subsubsection{Transaction Model Mismatch: Rigid ACID vs. Relaxed Semantics}

Traditional databases enforce a rigid one-size-fits-all transactional atomicity model. When this rigid semantics is applied blindly to metadata, it creates unnecessary bottlenecks.

\noindent 
\textbf{Single-shard hotspots.} Modern Machine Learning~(ML) workloads (e.g., Kubeflow~\cite{kubeflow}) frequently spawn tens of thousands of concurrent file creations within a single target directory. In a general-purpose database, every \create{} operation must synchronously update the shared parent-directory record, including attributes such as \texttt{num\_children} and \texttt{modify\_time}.
We quantify this contention penalty. Without contention, the previous system processed $\sim$60K\,QPS for \create{}. However, when all requests target the \textit{same} directory, throughput drops to $\sim$6K\,QPS (a 10$\times$ drop).

\noindent
\textbf{Unnecessarily strict atomicity for auxiliary metadata.} Updates to primary metadata~(see \autoref{fig:meta-table}) often trigger cascading updates to secondary indexes and user quotas~(auxiliary metadata). General-purpose databases execute these synchronously within transactions, forcing distributed two-phase commit~(2PC) since indexes/quotas reside on separate shards. In reality, file system semantics can tolerate minutes of latency for quota and index updates. In our production deployment, quota records are used for post-paid billing and monitoring with hourly settlement, so \emph{several minutes of propagation delay} between a primary metadata update and the corresponding quota update is acceptable. 
Thus, enforcing strict ACID guarantees here is an expensive over-design.

\subsubsection{Storage Implementation Mismatch: Long-Lived MVCC vs. Transient Metadata}

\begin{figure}

\begin{minipage}{0.23\textwidth}
    \centering
    \includegraphics[width=\linewidth]{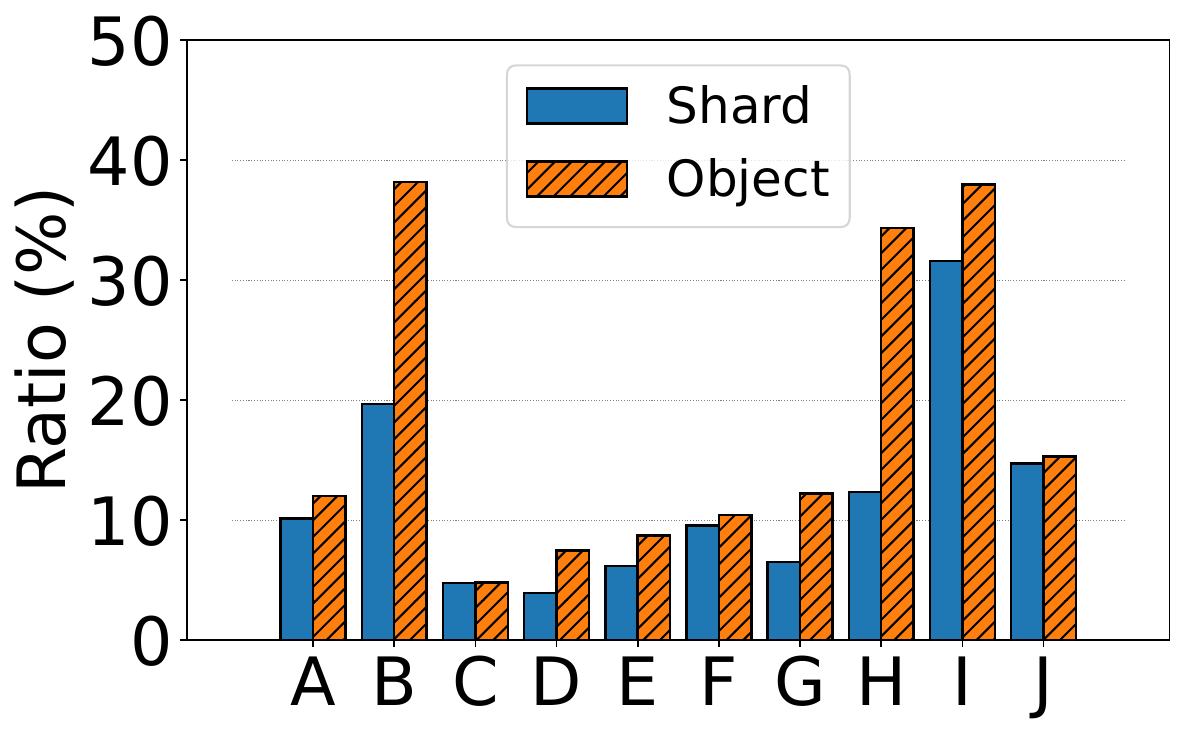}
    \caption{Ratio of objects with creation-time prefixes of 10 users~(from A to J).}
    \label{fig:motivation3-1}
\end{minipage}
\hfill
\begin{minipage}{0.23\textwidth}
    \centering
    \includegraphics[width=\linewidth]{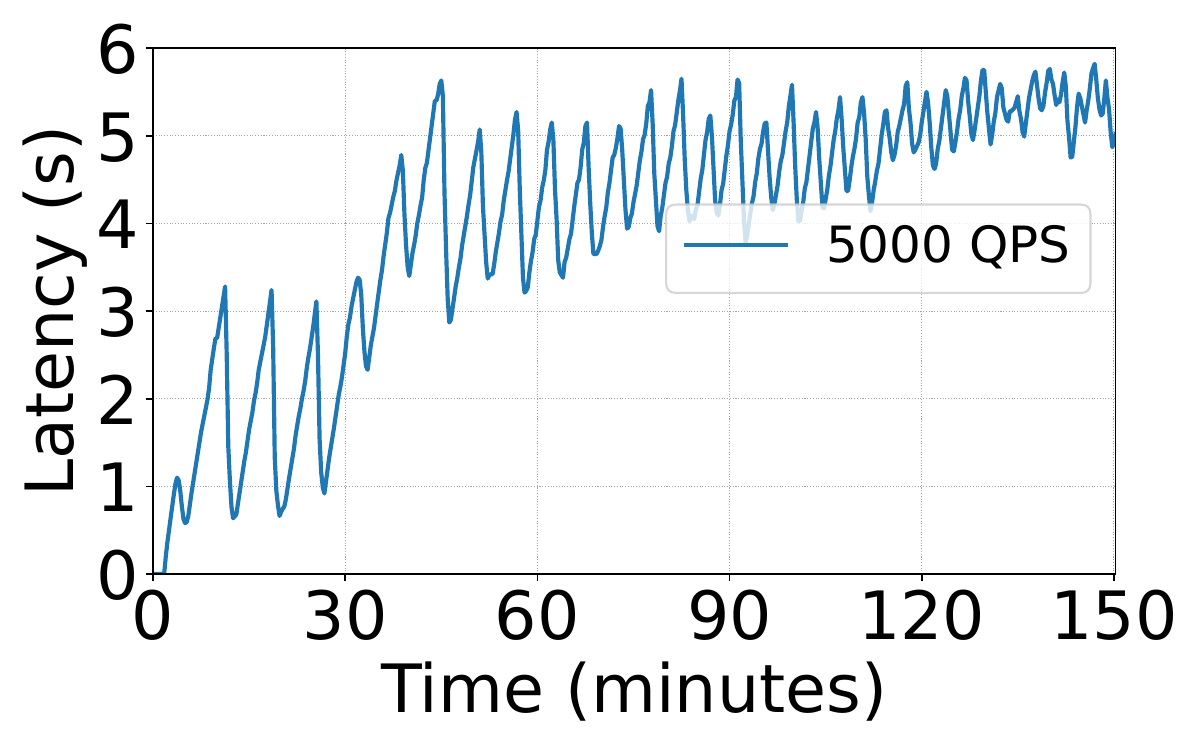}
    \caption{The effect of consecutive bulk deletion on \scan{} latency.}
    \label{fig:motivation-3-2}
\end{minipage}

\end{figure}

General-purpose storage engines~(e.g., LSM-trees with MVCC) are designed for arbitrary transactional workloads, including potentially long-running transactions and queries, making them poorly suited to the metadata lifecycle.

\noindent
\textbf{Metadata exhibits massive consecutive bulk deletions.} Object storage environments see frequent, large-scale range deletions driven by Time-To-Live (TTL)~expirations or explicit bucket purges~\cite{amazon-s3}. 
\autoref{fig:motivation3-1} shows the ratio of shards with creation-time prefixes for object names, and the ratio of objects named with creation-time prefixes.
This shows that many objects in our \bos{} clusters use time-based naming prefixes, which turns TTL expirations into dense, spatially continuous deletions in the database.

\noindent
\textbf{Transient versions accumulate as tombstones.} 
Although metadata transactions are exceptionally short~($<10$\,ms), traditional MVCC implementations preemptively flush multi-version states to disk to support potentially long-running queries, resulting in millions of unnecessary MVCC tombstones during bulk deletions.
When foreground range scans~(e.g., \scan{}) traverse these accumulated tombstones in the LSM-tree, they suffer severe read amplification. As demonstrated in \autoref{fig:motivation-3-2}, the latency of a 1,000-item \scan{} spikes from $<100$\,ms to $\sim$6 seconds during a deletion phase. Furthermore, because these deletions are highly localized, the database's generic full-shard compaction mechanism becomes too slow and resource-intensive to purge these tombstones in a timely manner.
\section{\metadb{} Design}
\label{sec:tafdb-design}

\subsection{Data Model}
\label{sec:design-records}

\begin{figure}[t!]
    \centering
    \includegraphics[width=1\linewidth]{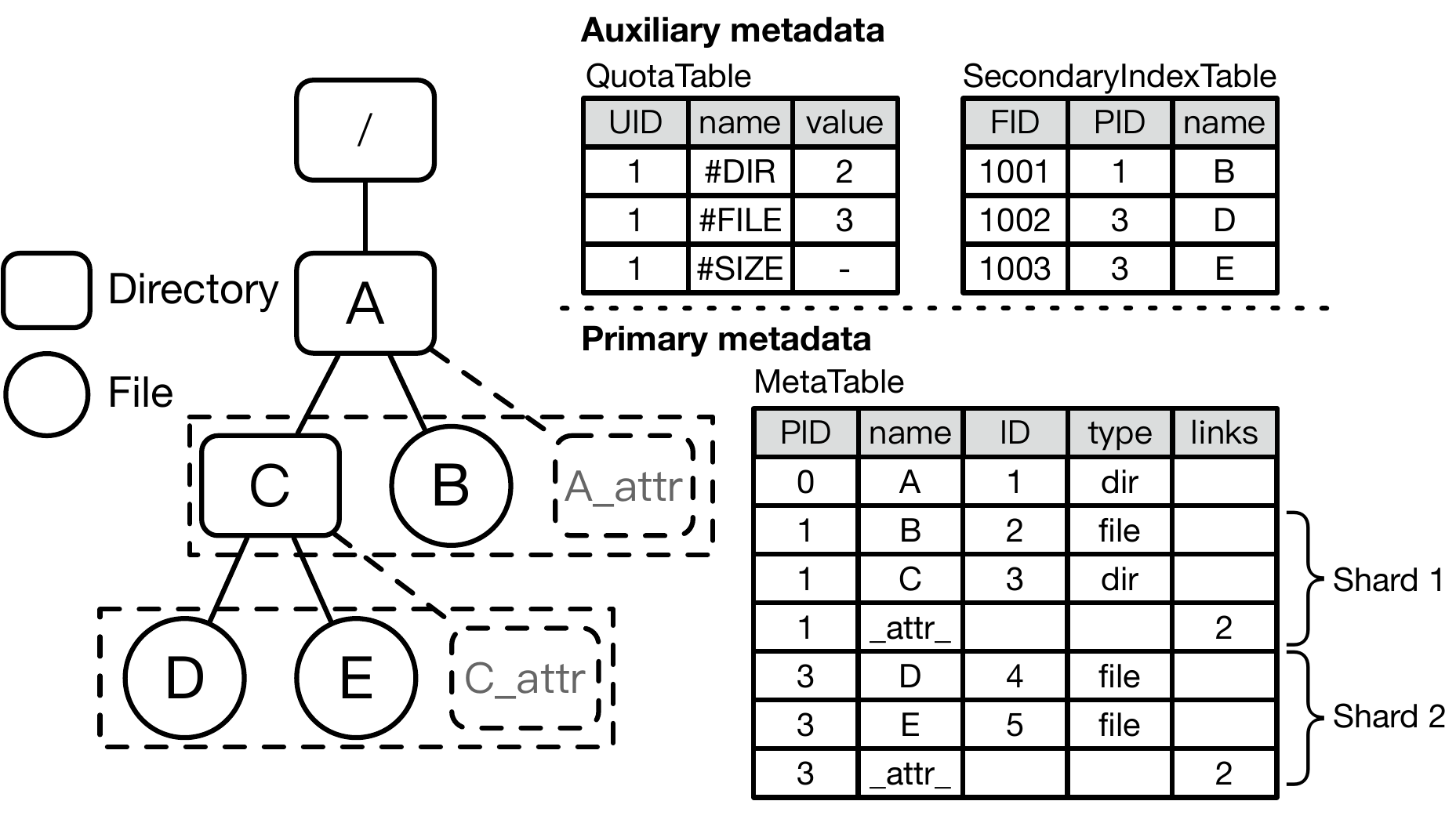}
    \caption{Metadata schema in \metadb{}. 
    In these tables, ID (identification) is the `inode ID' of each file or directory, PID is the ID of the entry’s parent directory in the directory hierarchy, UID is the user ID for quota accounting, and FID is the ID of the file in the data plane~(usually a separate system).}
    \label{fig:meta-table} 
\end{figure}

\metadb{} adopts a range-partitioned data model in which records are sorted and stored by primary key. Three tables are defined in \metadb{} as shown in \autoref{fig:meta-table}. MetaTable stores \emph{primary metadata} while QuotaTable and SecondaryIndexTable store \emph{auxiliary metadata}.

To support hierarchical namespaces, \textit{MetaTable} maps the directory tree into a flat schema. Following prior file systems such as InfiniFS~\cite{fast22-infinifs, eurosys23-cfs}, \metadb{} separates \emph{access} fields from \emph{attributes}. 
The access fields $(PID, name, ID)$ encode the directory hierarchy, while the remaining columns store attributes such as type, link count, and the number of children. 
In particular, the attributes of a directory are represented by a special entry named \texttt{\_attr\_}, whose parent is the directory that owns the attributes.
\textit{MetaTable} is keyed by $(PID, name)$, so a directory's \texttt{\_attr\_} entry is colocated with all of its children. 
We choose this layout because a single-directory operation typically updates both a child entry and the attributes of its parent directory. 
Placing them together preserves locality and, combined with our policy/size-based sharding~(\S\ref{sec:sharding-policy}), keeps common single-directory operations within a single shard.

Beyond \textit{MetaTable}, \metadb{} maintains two auxiliary tables. 
\textit{QuotaTable} tracks per-user resource usage for quota management. 
\textit{SecondaryIndexTable} stores mappings from files in the data plane to their corresponding metadata entries, providing an auxiliary index for lookup and debugging.

\subsection{Architecture}
\label{sec:design-arch}

\begin{figure}[t!]
    \centering
    \includegraphics[width=0.9\linewidth]{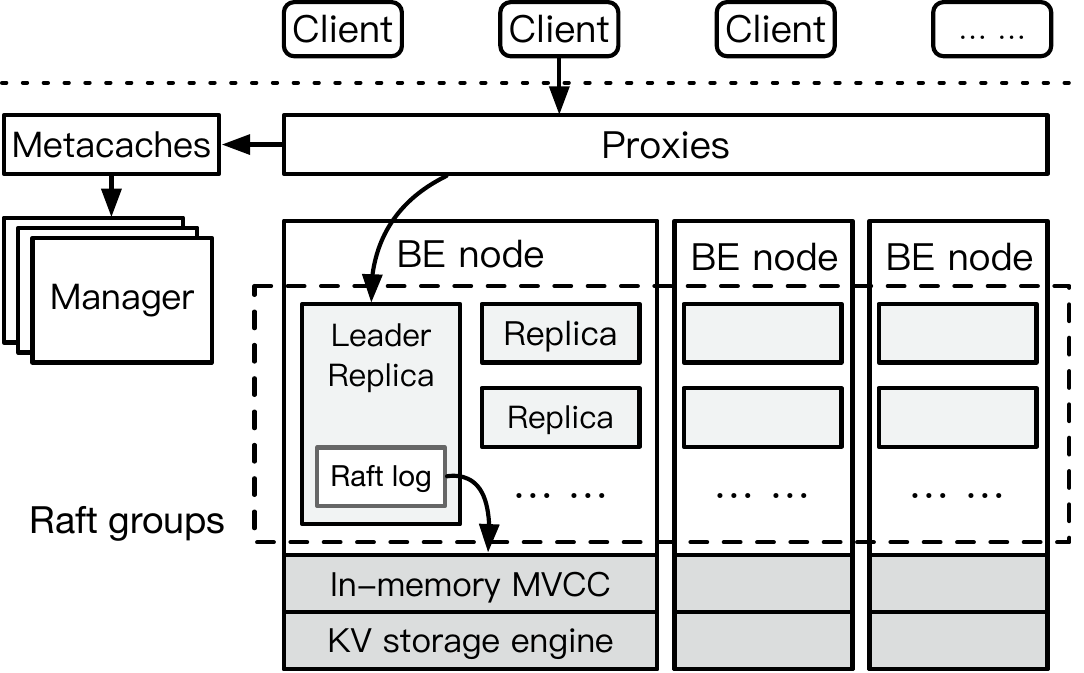}   
\caption{The architecture of a \metadb{} cluster. Proxies handle client routing and SQL parsing. The Manager oversees cluster status and dynamic sharding. MetaCache provides high-speed routing resolution. Backend (BE) nodes manage data replication via Raft and persist data using an integrated in-memory MVCC and RocksDB storage engine.}    \label{fig:arch} 
\end{figure}

We describe the \metadb{} architecture (\autoref{fig:arch}) through the lifecycle of a request.
When a client issues a metadata operation, it connects to a randomly selected stateless Proxy for load balancing.
The Proxy parses the lightweight SQL request, extracts the routing keys, and queries the MetaCache to locate the target records.
MetaCache provides fast shard-routing and consults the central Manager only on a cache miss.
Once the target backend~(BE) node is identified, the Proxy forwards the request to the leader replica of the corresponding shard.

In the BE node, the request is first appended to the Raft log and replicated across a three-node Raft group to provide availability and durability~\cite{atc14-raft}. 
After consensus, the operation is applied to the local storage engine. 
For primary metadata workloads that require both point lookups and range scans, each BE node integrates an in-memory MVCC layer with a single persistent RocksDB~\cite{rocksdb} instance per disk. 
To avoid double-logging, we disable RocksDB's native Write-Ahead Log~(WAL) and rely solely on the Raft log for durability~\cite{fast22-double-logging, nsdi26-xll}. 
Raft log truncation is deferred until the corresponding data is safely persisted in the local single-node storage engine.

\subsection{Addressing the Mismatches}
\label{sec:design-mismatches}

To bridge the gap between traditional databases and metadata workloads, \metadb{} introduces three core design pillars, each directly targeting one mismatch identified in \S\ref{subsec:motivation}.

\textbf{\ding{172} Tree-Topology-Induced Sharding and Execution~(\S\ref{sec:tree-sharding}).}
To address the Data Model Mismatch, \metadb{} uses a policy/size-based sharding mechanism that aligns shard boundaries with the directory tree topology and keeps the entries of a directory on the same shard. 
In addition, \metadb{} pushes single-shard transactional logic into BE nodes to eliminate unnecessary network round-trips.

\textbf{\ding{173} Semantic-Aware Off-Critical-Path Processing~(\S\ref{sec:side-path}).}
To address the Transaction Model Mismatch, \metadb{} removes aggregatable directory attributes and auxiliary metadata updates from the transactional critical path. Using an asynchronous bypass mechanism and an in-memory queue, it safely delays quota/secondary-index updates and resolves lock contention without sacrificing correctness.

\textbf{\ding{174} Lifecycle-Aware Storage Engine~(\S\ref{sec:storage-engine}).}
To address the Storage Implementation Mismatch, \metadb{} introduces an in-memory MVCC layer that prevents MVCC tombstones from reaching disk during bulk deletions. It further employs locality-aware range compaction to purge dense tombstone accumulations timely and ensure stable range-scan latency.
\section{Tree-Topology-Induced Sharding and Execution}
\label{sec:tree-sharding}

\metadb{} co-designs the data placement and execution engine according to the directory tree topology. We introduce a policy/size-based sharding mechanism to confine single-directory operations to a single shard~(\S\ref{sec:sharding-policy}) and further push down the transactional execution logic directly into the storage nodes to eliminate network coordination overhead~(\S\ref{sec:cas}).

\subsection{Sharding Policy}
\label{sec:sharding-policy}

Production metadata workloads exhibit strong spatial locality: our analysis shows that over 96\% of directory-entry updates are single-directory operations~(\S\ref{sec:moti-op-dsitr}). Single-directory operations access only the target entry and the attributes of its immediate parent directory. However, traditional size-based sharding splits shards at fixed size thresholds without considering directory-tree topology, often separating a parent directory from its children and thereby turning single-directory updates into distributed transactions.

To preserve structural colocation, \metadb{} implements a \textbf{policy/size-based sharding} mechanism. We introduce a schema-level attribute named \texttt{group\_key}. When a shard exceeds the configured capacity threshold, \metadb{} does not split it arbitrarily. 
Instead, the system splits the data at the \texttt{group\_key} boundary. By using the parent directory identifier as the \texttt{group\_key}, \metadb{} explicitly anchors a directory's attributes and its direct children within the same physical database shard. This keeps high-frequency single-directory operations in a single shard, avoiding distributed transactions.

While structural colocation avoids distributed transactions for single-directory operations, it also creates jumbo shards: unsplittable shards caused by exceptionally large directories. These jumbo shards can become performance bottlenecks, so \metadb{} uses two techniques to handle them:
(1) \textit{Jumbo-shard-aware scheduling and isolation.} \metadb{} continuously monitors shard load to identify unsplittable jumbo shards. The scheduler migrates hot jumbo shards to under-utilized nodes and evicts unrelated replicas, eliminating resource contention and dedicating the full machine to the jumbo directory.
(2) \textit{Workload-adjustable dual-mode sharding.} By default, \metadb{} keeps a directory in one shard to avoid distributed transactions. 
If a jumbo shard still exceeds a single machine's capacity after isolation, operators can split it across shards manually, but updates to that directory may require distributed transactions. This serves as a fallback for extreme cases.

\subsection{Single-Shard Transaction Logic Pushdown}
\label{sec:cas}

Our topology-aware sharding confines single-directory operations to a single physical node. However, executing these localized operations through a traditional database interface still incurs substantial overhead because general-purpose databases use a coordinator-based transaction model. Metadata operations inherently follow a Read-Modify-Write~(RMW) pattern. For example, a \texttt{mkdir} must read the current parent-directory attributes, insert the new child entry, and write back the updated parent attributes. In a typical design, a stateless proxy or client coordinates this sequence of operations, requiring at least two network round-trips to fetch state and commit the modification.

To eliminate this overhead, \metadb{} pushes transaction logic down to the storage node. Because our sharding policy keeps a directory's entries and attributes colocated, \metadb{} can bypass the coordinator and directly execute the entire directory-tree operation on the BE node. The BE node performs the full RMW sequence locally, removing multi-round network coordination and thereby reducing network overhead to achieve ultra-low latency for single-directory operations.
\section{Semantic-Aware Off-Critical-Path Processing}
\label{sec:side-path}

General-purpose databases enforce a uniform and rigid ACID transaction model across all data fields. 
This choice creates performance bottlenecks for metadata workloads. 
Specifically, concurrent operations within the same directory result in concurrent updates to its attributes, leading to severe lock contention. 
Furthermore, maintaining derived auxiliary metadata in the same transaction with primary metadata updates results in expensive multi-shard transactions. 

To eliminate these bottlenecks, \metadb{} adopts a semantic-aware off-critical-path processing mechanism. 
By understanding the unique consistency requirements of different metadata fields, \metadb{} extracts both \emph{aggregatable directory attributes} and \emph{auxiliary metadata} from the transaction and performs operations on these fields through off-critical-path processing.

\subsection{Decoupled Asynchronous Auxiliary Updates}
\label{sec:async-index}

\begin{figure}[t!]
    \centering
    \includegraphics[width=1\linewidth]{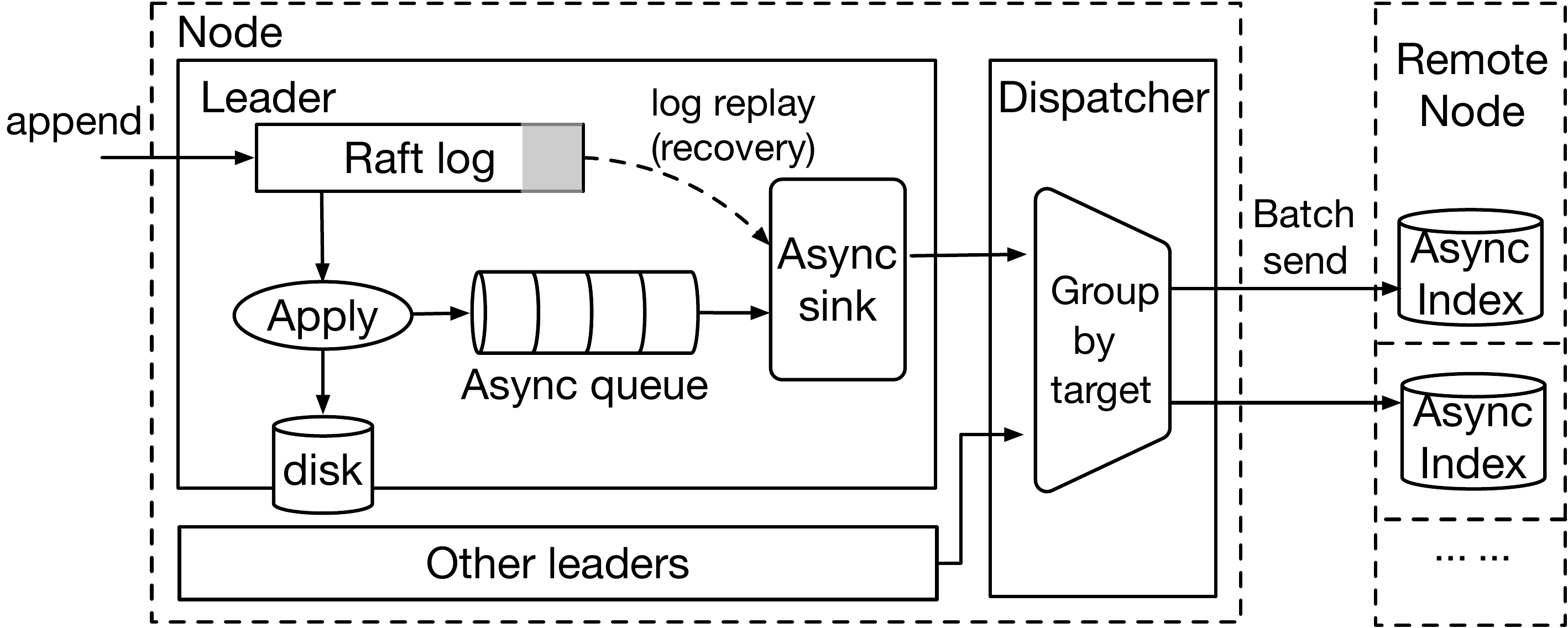}
    \caption{Asynchronous updating framework.}
    \label{fig:async-index}
\end{figure}

Metadata operations maintain global auxiliary metadata such as user quotas and secondary indexes. Because this metadata usually resides on different shards from the primary metadata, updating it atomically with the primary metadata requires multi-shard transactions, which reduces throughput.

Unlike primary metadata, this auxiliary metadata does not require atomicity with primary metadata updates. 
\metadb{} therefore decouples quota and index updates from the primary metadata update transaction and processes them through an asynchronous bypass mechanism. 
As shown in \autoref{fig:async-index}, \metadb{} generates auxiliary metadata update requests during the primary metadata updates, persists them in the local Raft log, applies the primary metadata update to the single-node storage engine, and then enqueues the auxiliary metadata update requests into an in-memory asynchronous queue. An asynchronous sink drains this queue and batches requests by destination before forwarding them to remote shards, reducing network overhead. 
This bypass removes quota and index updates from the critical path and avoids the multi-shard transactions that are otherwise required.

This design must handle two issues: lost updates and duplicated updates. To avoid lost updates, each auxiliary metadata update request is first persisted in the Raft log before being appended to the in-memory queue. The Raft log is truncated periodically based on successful sink progress, while requests not yet applied on the quota/index shard are retained so they can be replayed after failures. Upon failure or leader change, the new leader replays the Raft log after the latest checkpoint to reconstruct the in-memory queue and resume the sink. Because checkpoints advance in batches, recovery may replay requests that have already been applied. To suppress such duplicates, each auxiliary update request carries a timestamp, which is recorded in the quota/index table to ensure that the same request is applied at most once. These mechanisms ensure that asynchronous auxiliary updates are neither lost nor applied more than once.

\subsection{Conflict-Free Asynchronous Attribute Aggregation}
\label{sec:async-attr}

\begin{figure}[t!]
    \centering
    \includegraphics[width=1\linewidth]{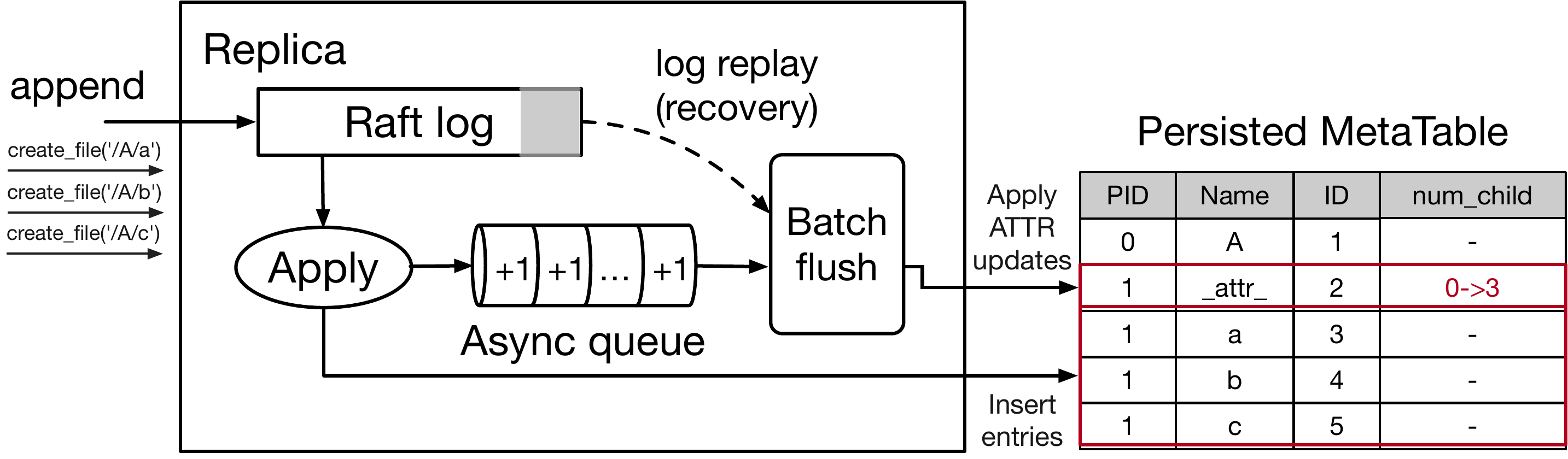}
    \caption{Asynchronous attribute update.}
    \label{fig:async-attr}
\end{figure}

Directory tree modifications usually need to update parent-directory attributes. For example, creating a file typically requires incrementing the \texttt{num\_children} field and updating the \texttt{modify\_time} field. A typical implementation often locks the parent-directory attribute record to ensure atomicity. Under high concurrency, a large number of modifications within the same directory contend on these shared attributes.

\metadb{} observes that these attribute updates (hotspots) have aggregatable semantics. For \texttt{num\_children}, the final value is the sum of all deltas. For \texttt{modify\_time}, the final value is the committed value with the largest timestamp. Thus, for concurrent modifications in the same directory, these attribute updates can be combined through aggregation. 

Based on this observation, \metadb{} no longer synchronously updates parent-directory attribute records within transactions. 
Instead, it moves the updates of these attribute records off the critical path and updates them asynchronously.
\metadb{} first persists attribute changes in the Raft log for durability, then places them into an in-memory queue, where a background thread asynchronously aggregates them and writes them to the storage engine in batches. 
Specifically, \texttt{num\_children} is aggregated by accumulating deltas, while \texttt{modify\_time} is determined by maintaining the value committed with the largest timestamp. This design avoids hotspot-attribute lock contention on the write path by removing attribute updates from the foreground transaction.

On the read path, before returning directory attributes to the client, the system merges persisted states with pending attribute changes in the in-memory queue, so that the client always reads the up-to-date values.
Merging pending attribute changes in memory adds negligible latency.
To ensure correctness, asynchronous attribute aggregation reuses the same asynchronous update framework with the same persistence guarantees as auxiliary metadata updates, ensuring that pending changes are neither lost nor applied more than once.

\section{Lifecycle-Aware Storage Engine}
\label{sec:storage-engine}

\begin{figure}[t!]
    \centering
    \includegraphics[width=0.9\linewidth]{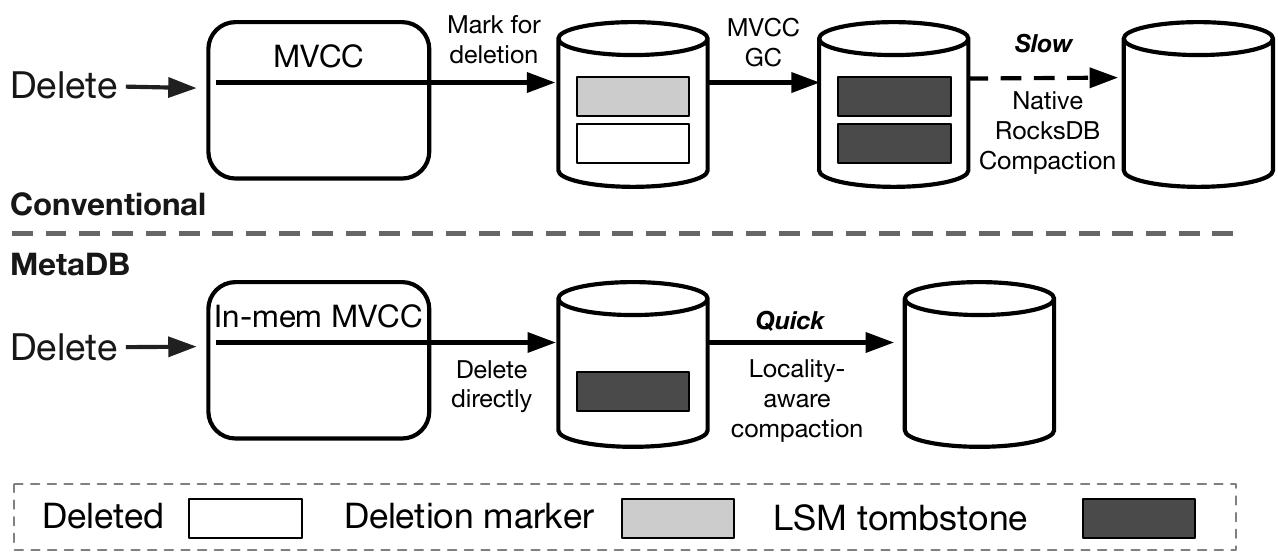}
    \caption{General architecture of preventing tombstones from accumulating in the LSM-tree-based storage engine.}  
    \label{fig:mvcc-all}  
\end{figure}

Our cloud object storage infrastructure experiences a massive volume of consecutive deletions driven by lifetime expiration policies and user-initiated purges. In a typical LSM-tree-based storage engine, these deletions trigger a rapid accumulation of tombstones, which degrades the latency of range scans like \scan{}. To resolve this Storage Implementation Mismatch, \metadb{} redesigns the storage layer specifically for the lifecycle of metadata as shown in~\autoref{fig:mvcc-all}. We introduce an in-memory MVCC mechanism to eliminate the persistence of redundant MVCC tombstones~(\S\ref{sec:in-memory-mvcc}) and a locality-aware compaction strategy to clean up LSM-tree tombstones~(\S\ref{sec:locality-aware-compaction}).

\subsection{In-Memory MVCC}
\label{sec:in-memory-mvcc}

General-purpose distributed databases typically employ MVCC for concurrency control and LSM-trees for persistent storage~\cite{tidb, cockroachdb}. Under this architecture, a deletion operation first creates a logical MVCC tombstone, and the underlying LSM-tree then performs the physical deletion by writing its own tombstone. Because traditional databases must support long-running transactions, they routinely persist MVCC tombstones to disk. This creates a double-tombstone problem in which both MVCC and LSM-tree tombstones accumulate, significantly increasing read amplification.

Because metadata transactions typically finish in a few milliseconds, \metadb{} keeps multiple committed versions and MVCC tombstones only in memory. During a flush, the system prunes the in-memory multi-version state and persists only the latest committed state of each key to the persistent key-value store. For normal updates, this means writing only the latest committed version while discarding stale versions. For deletions, \metadb{} does not persist the in-memory MVCC tombstone itself. Instead, it represents the deletion directly by committing an LSM-tree tombstone to persistent storage. As a result, MVCC tombstones do not accumulate in persistent storage, thereby eliminating the double-tombstone penalty and significantly reducing read amplification.

Before flushing an in-memory table, \metadb{} retains it slightly longer than the 5-second transaction timeout, so that all transactions that may still read its versions have completed. If a node fails before the flush completes, the committed updates can be recovered by replaying the Raft log, whose entries are retained until the corresponding data have been durably persisted in the storage engine.

\subsection{Locality-Aware Compaction}
\label{sec:locality-aware-compaction}

While in-memory MVCC eliminates redundant multi-version data, the underlying LSM-tree still generates one tombstone per deletion. Our production metadata workloads contain massive consecutive bulk deletions. These bulk operations generate long runs of consecutive tombstones that severely penalize the performance of range scans like \scan{}.

Common LSM-tree compaction mechanisms are ill-equipped to handle this pattern. Native RocksDB compaction is triggered passively by level saturation. This delays tombstone cleanup and allows stale records to linger indefinitely~\cite{sigmod20-lethe}. Alternatively, forcing a full-shard compaction resolves the read amplification but incurs an unacceptable performance overhead for large database shards.

To resolve this, we observe that consecutive metadata deletions exhibit strong spatial locality. 
Thus, instead of passively waiting for global thresholds or aggressively compacting entire shards, \metadb{} dynamically tracks the specific key ranges experiencing heavy deletions. When the volume of deletions within a narrow tracked range surpasses a defined threshold, \metadb{} proactively triggers a localized compaction strictly confined to that affected area. This targeted approach efficiently purges accumulated tombstones with minimal resource overhead. By cleaning up consecutive tombstones timely, \metadb{} guarantees stable and predictable latency for range scans like \scan{}.
\section{Evaluation}
\label{sec:eval}

\subsection{Experimental Setup}
\label{sec:eval-setup}

\noindent
In the experiments, we use the same type of machines as our production cluster.
Our test cluster is configured with 4 BE nodes, 3 Manager nodes, and 3 Proxy nodes. 
Each node is configured with two Intel Xeon Platinum 8350C processors~(64 cores), 256 GB of DDR4-2933 memory, one 480\,GB SATA SSD, one 4\,TB NVMe SSD, and two 25\,Gbps network connections.
Each node runs CentOS 6.3 (Kernel 4.14.0).

To comprehensively evaluate the performance benefits of \metadb{}, we conduct a series of experiments using both synthetic microbenchmarks and real-world production traces. First, we separately examine the gains from our three core design pillars. Specifically, \S\ref{sec:expr-tree-structure} evaluates the effectiveness of our tree-topology-induced sharding and transaction pushdown for single-directory operations. \S\ref{sec:expr-async-update} shows how the semantic-aware off-critical-path processing improves system throughput under severe lock contention. \S\ref{sec:expr-lifecycle} validates the lifecycle-aware storage engine and shows that it sustains stable, low latency for range scans during massive consecutive bulk deletions. Next, \S\ref{sec:eval-stress-test} evaluates the performance of \metadb{} under stress tests in production environments.
Finally, \S\ref{sec:eval-production} shows nine representative production deployments, ratio of distributed transactions in two \cfs{} and two \mantle{} clusters, tombstone dynamics on a production node, and production operation latency.
We also compare the performance of \metadb{} and \texttt{TiDB} in \autoref{a1:eval-gp-db}.

\subsection{Effectiveness of Tree-Topology-Induced Sharding and Execution}
\label{sec:expr-tree-structure}

\begin{figure}[t]
\centering
    \centering
    \includegraphics[width=0.9\linewidth]{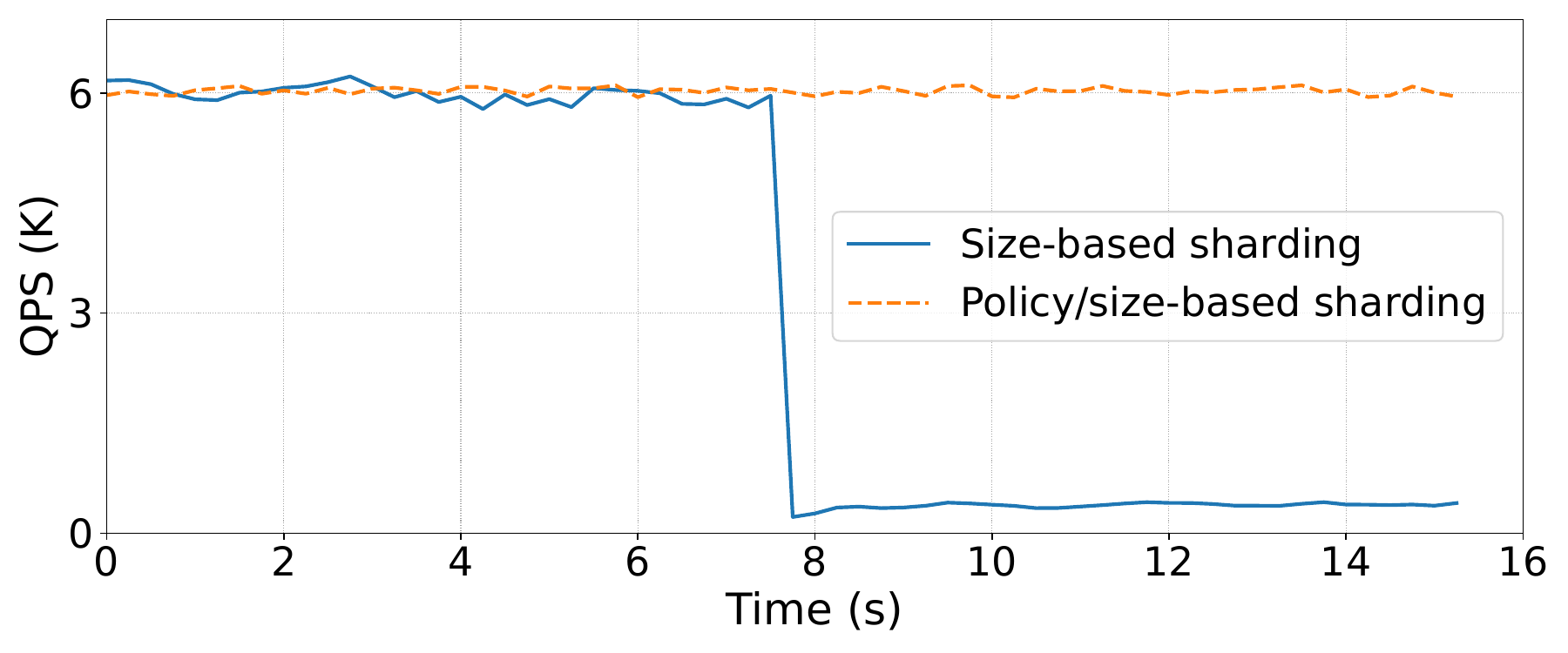}
    \caption{Throughput with \create{} under the same directory.}
    \label{fig:eval-sharding-throughput}
\end{figure}

\noindent
\textbf{Effectiveness of policy/size-based sharding.}
In this evaluation, we start with an empty directory, continuously create new files under it using four concurrent \create{} operations, and record the throughput.
\autoref{fig:eval-sharding-throughput} compares the throughput of \create{} under the same directory between size-based sharding and policy/size-based sharding. 
With size-based sharding, throughput drops from 6.0K\,QPS to 0.4K\,QPS after a split separates entries from the same directory. 
In contrast, policy/size-based sharding sustains 6.0K\,QPS throughout the run. 
This result shows that aligning shard boundaries with directory boundaries prevents common-case same-directory updates from turning into distributed transactions.

\begin{figure}[t!]
    \centering
    \begin{subfigure}[b]{0.48\linewidth}
        \includegraphics[width=\linewidth]{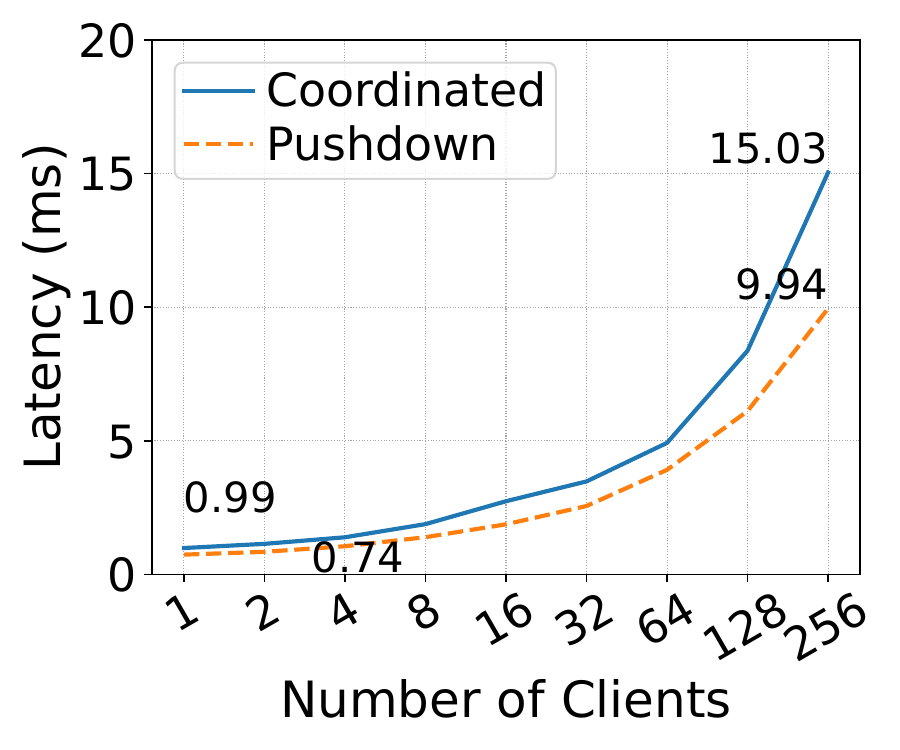}
        \caption{Latency}     
        \label{fig:eval1-cas-lat}
    \end{subfigure}
    \begin{subfigure}[b]{0.48\linewidth}
        \includegraphics[width=\linewidth]{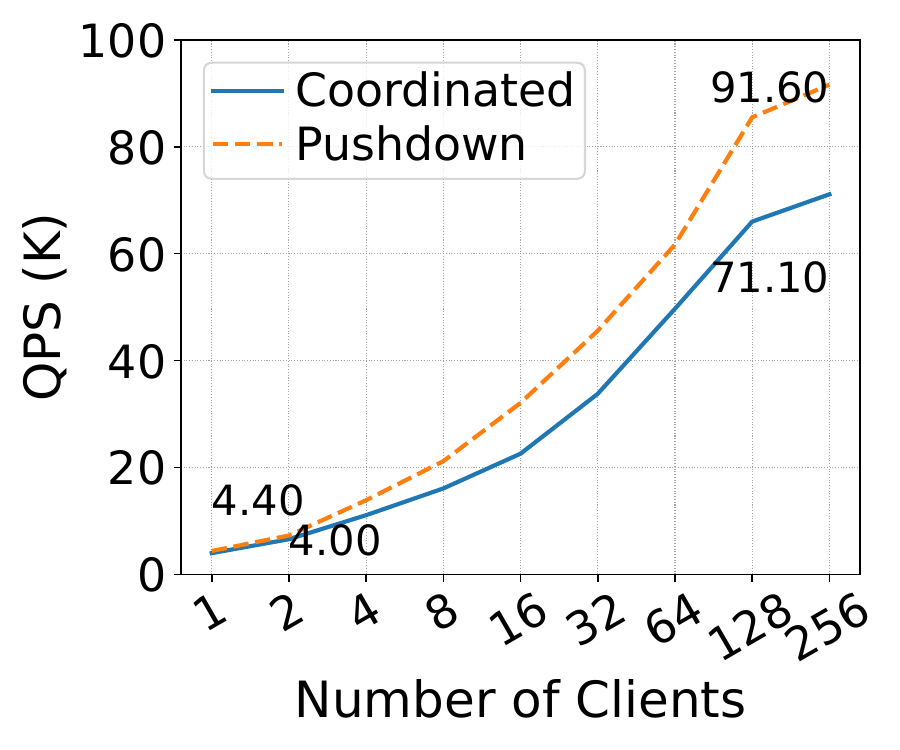}
        \caption{Throughput}
        \label{sec:eval1-cas-qps}
    \end{subfigure}
    \caption{Latency and throughput between coordinator-based execution and pushdown for \create{}.}
    \label{fig:eval-cas}
\end{figure}

\noindent
\textbf{Latency and throughput effect of single-shard transaction logic pushdown.}
We evaluate file creation using \create{} under varying client loads, comparing coordinator-based execution with single-shard transaction logic pushdown.
Each client has four threads where each thread creates files in a standalone directory to avoid lock contention on the directory's attributes~(\S\ref{subsec:motivation}).
The average latency and throughput are shown in \autoref{fig:eval-cas}.
With a single client, logic pushdown achieves 25.3\% lower latency and 10.0\% higher throughput than coordinator-based execution, while with 256 clients, it achieves 33.9\% lower latency and 28.8\% higher throughput, respectively.
This shows that single-shard transaction logic pushdown can reduce latency and increase throughput.

\subsection{Effectiveness of Semantic-Aware Off-Critical-Path Processing}
\label{sec:expr-async-update}

\begin{figure}[t]
\centering

\begin{minipage}{0.22\textwidth}
    \centering
    \includegraphics[width=\linewidth]{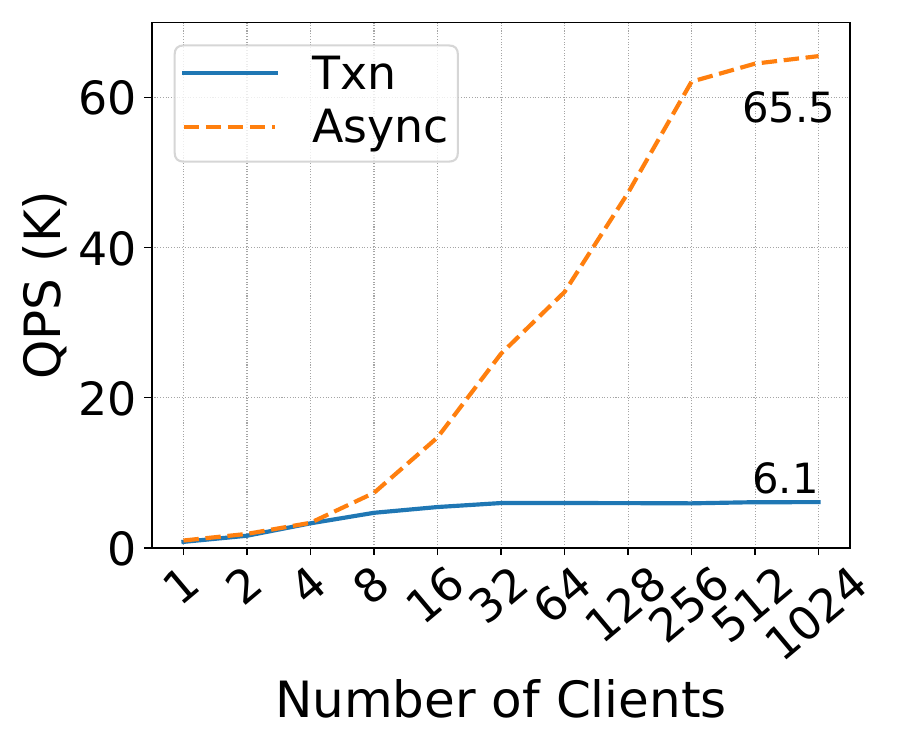}
    \caption{Throughput of lock-based transactional operation execution and conflict-free asynchronous attribute aggregation.}
    \label{fig:eval-2-async-attr}
\end{minipage}
\hfill
\begin{minipage}{0.22\textwidth}
    \centering
    \includegraphics[width=\linewidth]{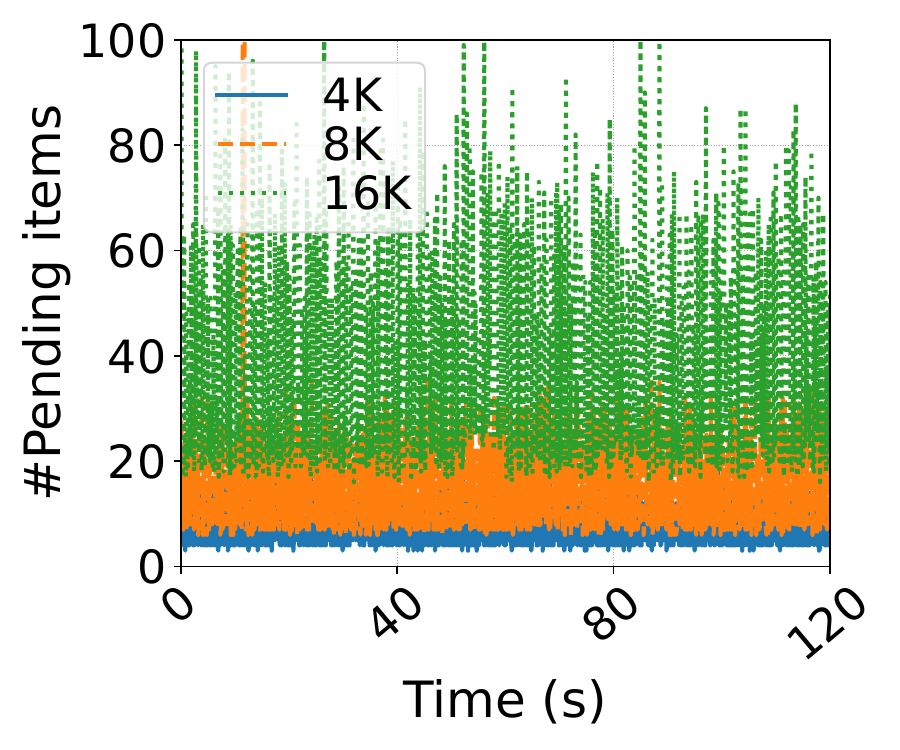}
    \caption{Number of pending quota updates in the asynchronous queue with asynchronous auxiliary updates over 120\,s.}
    \label{fig:quota-update-latency}
\end{minipage}
\end{figure}

\noindent
\textbf{Effectiveness of conflict-free asynchronous attribute aggregation.}
In this experiment, we issue \mkdir{} under the same directory with an increasing number of clients.
\autoref{fig:eval-2-async-attr} compares the throughput of lock-based transactional operation execution~(\texttt{Txn}) and conflict-free asynchronous attribute aggregation~(\texttt{Async}) with concurrent \mkdir{} under the same directory where each client sends four concurrent requests.
\texttt{Async} achieves 10.7$\times$ the throughput of \texttt{Txn}, demonstrating that our asynchronous attribute aggregation can achieve higher throughput than lock-based transactional execution under high concurrency of operations in the same directory by avoiding contention on directory attributes.

\noindent
\textbf{Latency of asynchronous updates.}
To quantify the lag between metadata updates and quota/index updates, we measure the number of pending quota updates at different \create{} throughputs.
\autoref{fig:quota-update-latency} shows the number of pending items in the asynchronous queue over a 120-second window at three throughput levels.
The number of pending items increases with operation throughput, while the peak remains below 100. 
This indicates a latency of less than 10\,ms at 16K\,QPS between the primary metadata being updated and the asynchronous updates being successfully sent to the remote shard.
As the index and quota updates can tolerate minutes of latency after the primary metadata is updated, the latency cost of the asynchronous framework is acceptable even with high concurrency of metadata operations.

\subsection{Effectiveness of Lifecycle-Aware Storage Engine}
\label{sec:expr-lifecycle}

\begin{figure}[t!]
    \centering
    \includegraphics[width=1\linewidth]{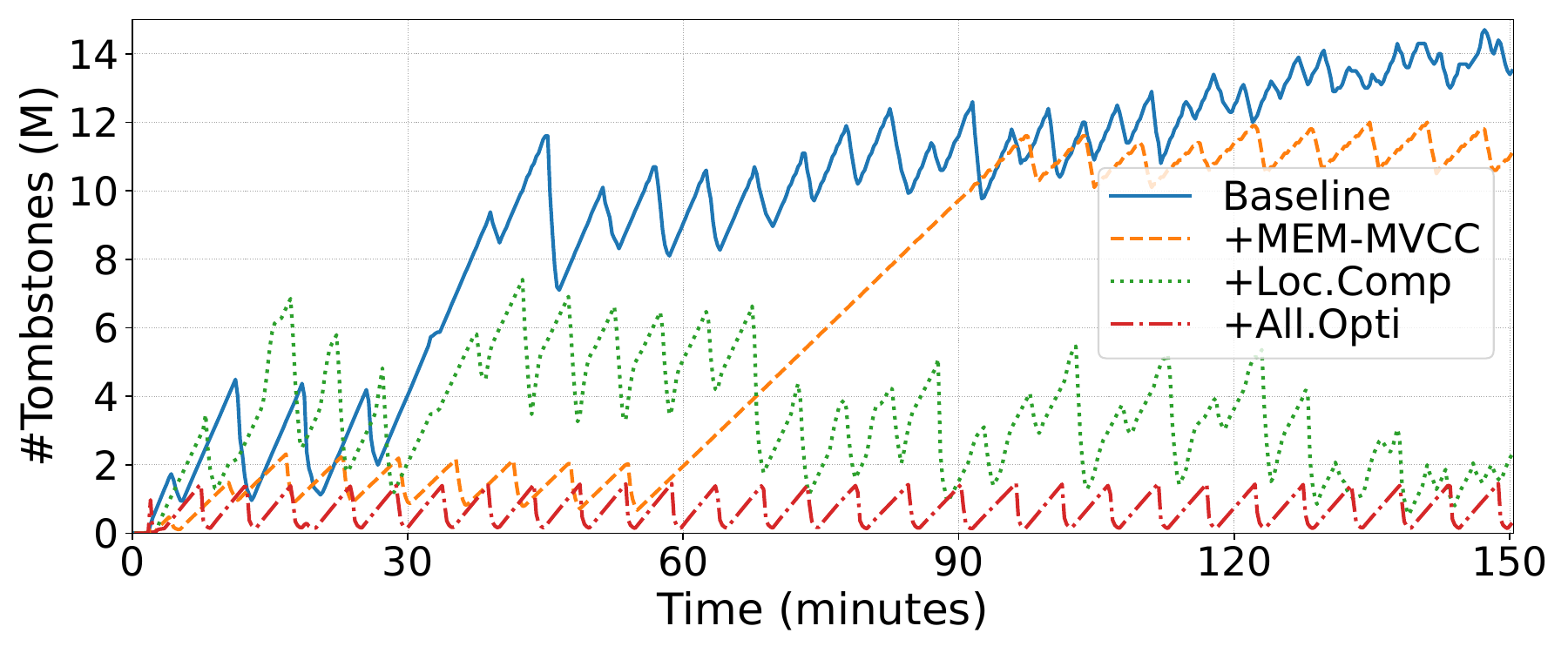}
    \caption{Number of tombstones.}   
    \label{fig:eval-deletion-tombstone}  
\end{figure}

\begin{figure}[t!]
    \centering
    \includegraphics[width=1\linewidth]{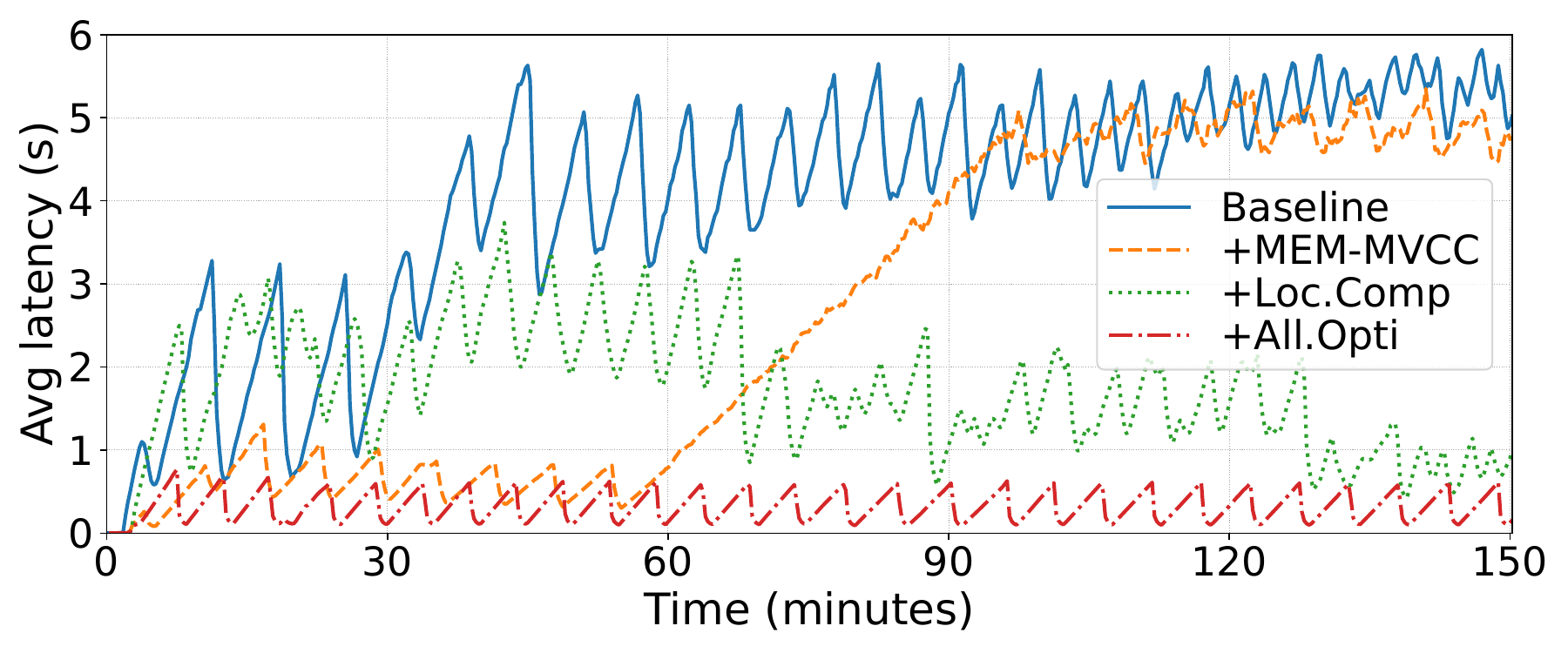}
    \caption{The effect on scan latency of bulk deletion.}   
    \label{fig:eval-deletion-scan-lat}  
\end{figure}

\begin{figure}[t!]

\begin{minipage}{0.23\textwidth}
    \centering
    \includegraphics[width=\linewidth]{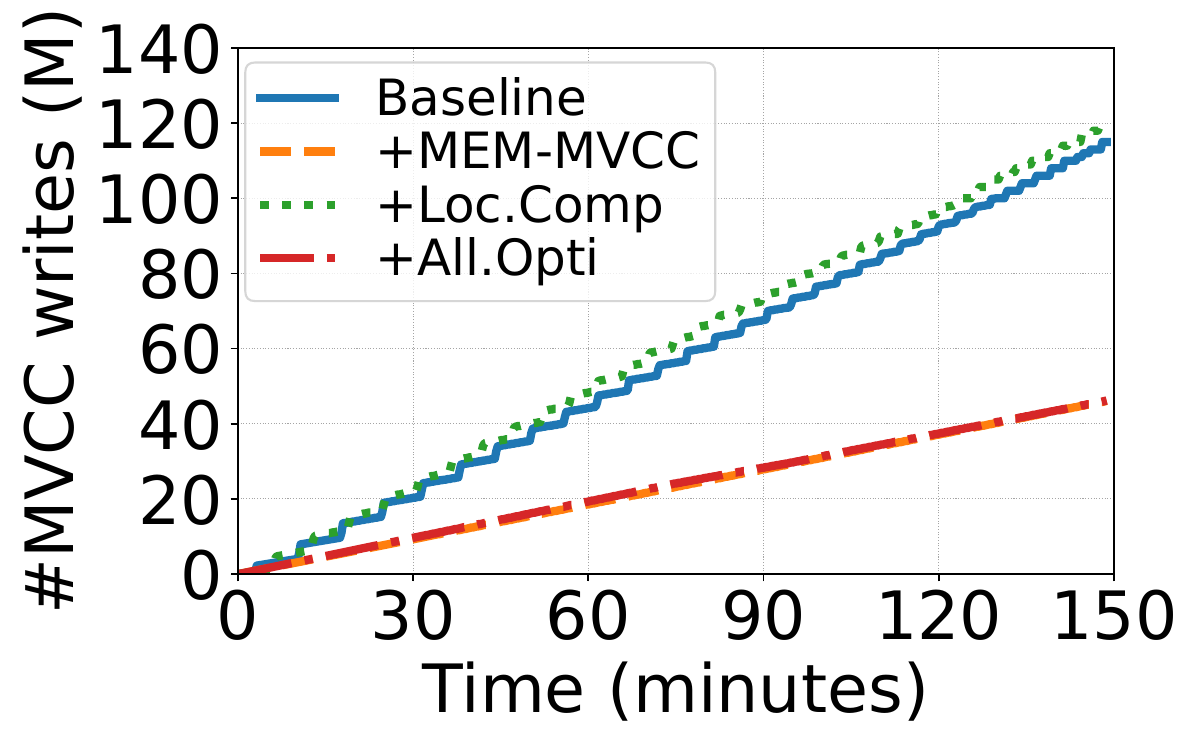}
    \caption{Number of MVCC writes.}
    \label{fig:eval-deletion-MVCC}
\end{minipage}
\hfill
\begin{minipage}{0.23\textwidth}
    \centering
    \includegraphics[width=\linewidth]{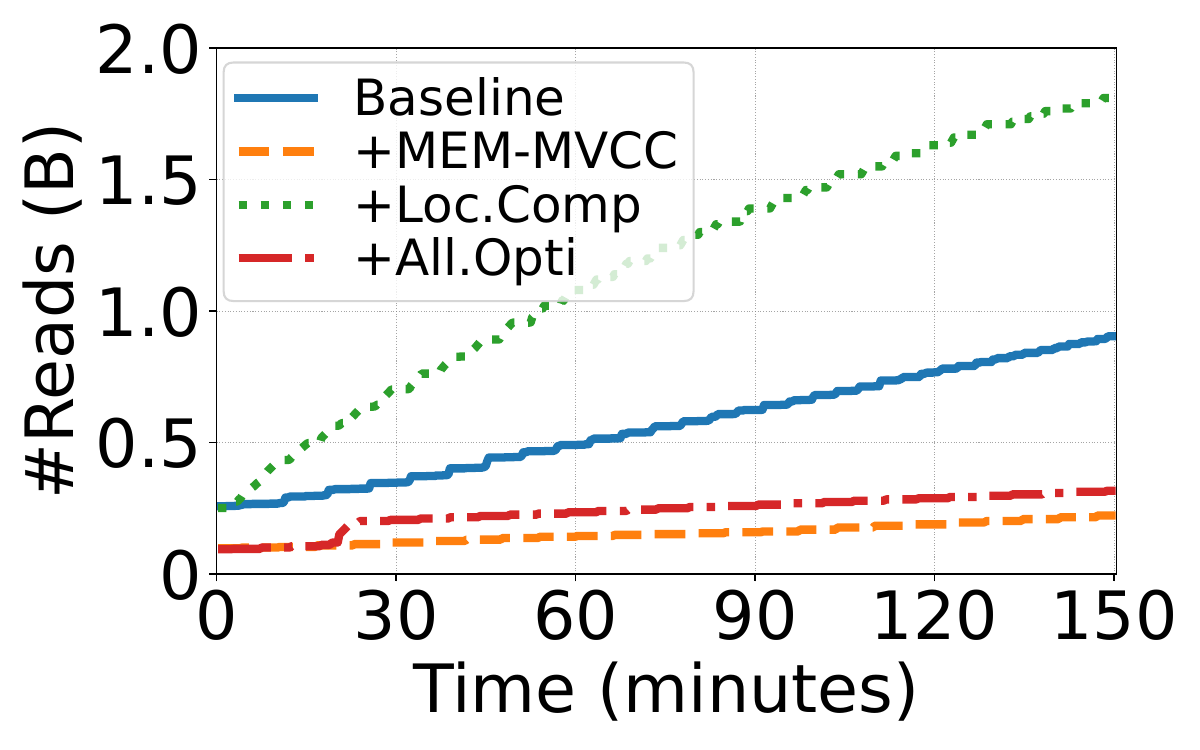}
    \caption{Number of compaction reads.}
    \label{fig:eval-deletion-compaction}
\end{minipage}
\end{figure}

To show how effectively in-memory MVCC reduces tombstone accumulation, we evaluate four configurations:
\ding{182} \emph{Baseline}: storage-backed MVCC with global compaction;
\ding{183} \emph{MEM-MVCC}: in-memory MVCC with global compaction;
\ding{184} \emph{Loc.Comp}: storage-backed MVCC with locality-aware compaction; and
\ding{185} \emph{All.Opti}: in-memory MVCC with locality-aware compaction.
We issue bulk deletions at 5K\,QPS while concurrently running a single \scan{} that returns 1K records. 
We record the number of tombstones in the storage engine, \scan{} latency, the number of KVs written to the storage engine, and the number of reads during compaction.

\noindent
\textbf{Tombstone accumulation.}
\autoref{fig:eval-deletion-tombstone} shows the number of tombstones in the 150-minute experiment.
\textit{Baseline} accumulates tombstones rapidly, reaching 14M and continuing to grow as deletions proceed.
With \textit{MEM-MVCC}, the number of tombstones grows more slowly than \textit{Baseline}. However, the count eventually reaches a similar level and fluctuates around 10M--12M, showing that RocksDB's global compaction cannot clean up tombstones quickly enough since each compaction merges only two adjacent levels, leaving tombstones in the remaining levels untouched.
\textit{Loc.Comp} removes tombstones more frequently and completely, so it controls the tombstone count better than the previous two configurations, but the count still fluctuates between 1M and 8M.
\textit{All.Opti} achieves the most stable behavior.
The number of tombstones remains consistently low, staying below 2M throughout the experiment, demonstrating the effectiveness of our design in preventing tombstone accumulation.

\noindent
\textbf{Effect on \scan{} latency.}
\autoref{fig:eval-deletion-scan-lat} shows the \scan{} latency under the same deletion workload as in \autoref{fig:eval-deletion-tombstone}.
Across all configurations, the \scan{} latency closely follows the tombstone accumulation trend.
For both the \textit{Baseline} and \textit{MEM-MVCC}, the latency is higher and less stable than with the two locality-aware compaction configurations~(\textit{Loc.Comp} and \textit{All.Opti}) as the number of tombstones accumulates, fluctuating around 5\,s after sustained deletions.
With \textit{Loc.Comp}, \scan{} latency fluctuates between 1\,s and 3\,s.
Enabling both \textit{MEM-MVCC} and \textit{Loc.Comp}~(All.Opti) delivers the lowest and most stable latency, peaking at 0.7\,s, matching the reduced tombstone counts observed earlier.
This shows that our design can provide low and stable \scan{} latency by preventing tombstones from accumulating in LSM-tree-based storage.

\noindent
\textbf{Reduction in KV writes with in-memory MVCC.}
\autoref{fig:eval-deletion-MVCC} shows the number of writes to the LSM-tree during the experiment, where writes grow as the total number of deletions increases over time.
With in-memory MVCC enabled~(\textit{MEM-MVCC} and \textit{All.Opti}), the number of writes is about 60\% lower than with storage-persisted MVCC~(\textit{Baseline} and \textit{Loc.Comp}), showing that in-memory MVCC effectively reduces write volume, reducing write amplification.

\noindent
\textbf{Impact on compaction reads.}
\autoref{fig:eval-deletion-compaction} shows the number of reads from the LSM-tree during compaction.
\textit{Loc.Comp} has the highest number of reads since compaction is more frequently triggered in this configuration.
With in-memory MVCC~(\textit{MEM-MVCC} and \textit{All.Opti}), the number of reads is lower than that under the \textit{Baseline} and \textit{Loc.Comp}, owing to the reduced amount of data written to the LSM-tree.

\subsection{Overall Performance Stress Testing}
\label{sec:eval-stress-test}

\begin{figure}[t!]
    \centering
    \begin{subfigure}[b]{0.48\linewidth}
        \includegraphics[width=\linewidth]{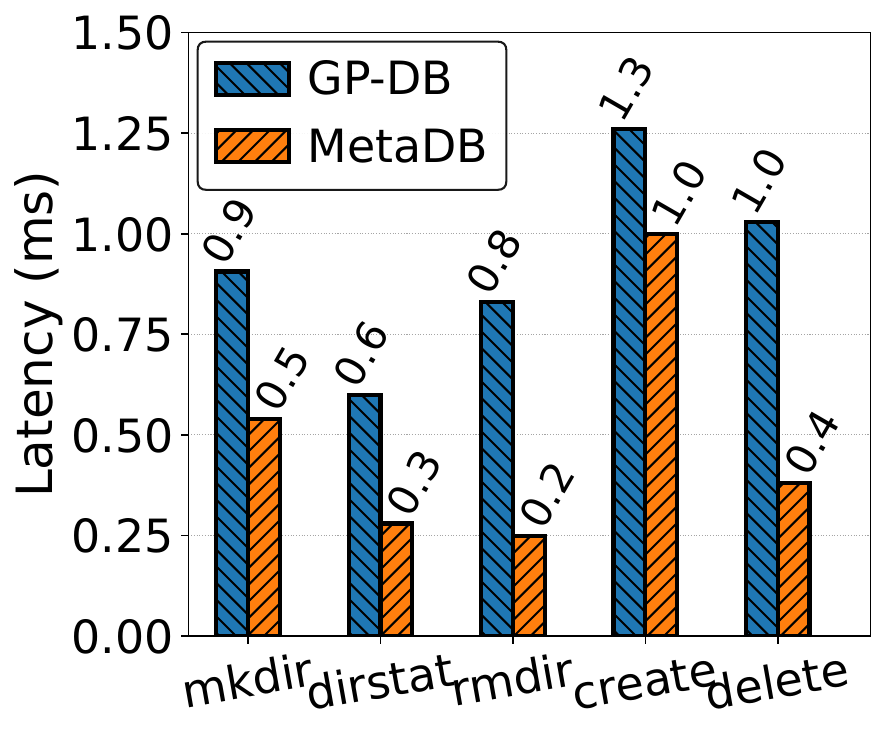}
        \caption{Latency}     
        \label{fig:eval4-prod-lat}
    \end{subfigure}
    \begin{subfigure}[b]{0.48\linewidth}
        \includegraphics[width=\linewidth]{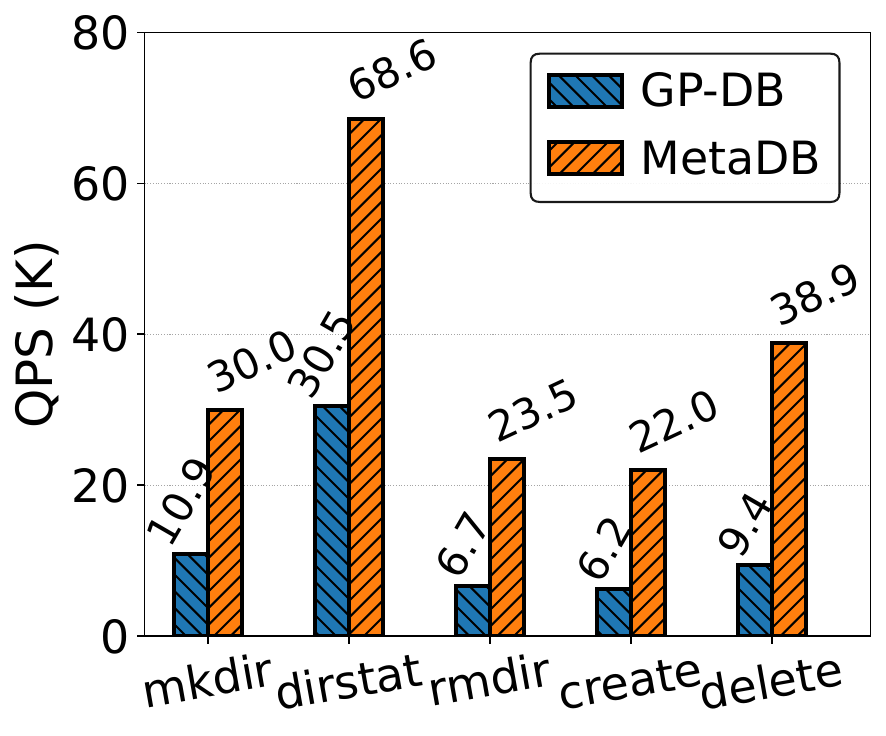}
        \caption{Throughput}
        \label{sec:eval4-prod-thr}
    \end{subfigure}
    \caption{Latency and throughput between general-purpose databases and \metadb{} on our production cluster.}
    \label{fig:eval-4-prod-test}
\end{figure}

We measure the latency and throughput in one production cluster with 10 BE nodes and issue metadata operations with \texttt{mdtest}, where \texttt{mdtest} is treated as a user in the corresponding cluster.
For each test, we start 20 clients, each with 50 threads sending requests concurrently.
Each client sends 2,000 requests in total.
\autoref{fig:eval-4-prod-test} compares the latency and throughput of five metadata operations between the general-purpose database~(GP-DB) and~\metadb{}.
Compared with GP-DB, \metadb{} reduces latency by 23.1\%--75.0\% and achieves 2.2$\times$--4.1$\times$ its throughput.

\subsection{\metadb{} in Production}
\label{sec:eval-production}

\begin{table}[t!]
    \footnotesize
    \centering   
    \caption{Nine representative \metadb{} deployments.}  
    \label{table:production}
    \begin{tabular}{l|r|r|r|r|r|r}
    \toprule
    \textbf{\textit{ID}} &  \makecell[r]{\textit{\textbf{Num}}
    \\ \textit{\textbf{Nodes}}} & \makecell[r]{\textit{\textbf{\#Rec}}
    \\ \textit{\textbf{\,(B)}}} & \makecell[r]{\textit{\textbf{Max\,R}}
    \\ \textit{\textbf{\,QPS(K)}}} & \makecell[r]{\textit{\textbf{Max\,W}}
    \\ \textit{\textbf{\,QPS(K)}}} & \makecell[r]{\textit{\textbf{Data}} \\ \textit{\textbf{(PB)}}} & \makecell[r]{\textit{\textbf{Index}} \\ \textit{\textbf{gap}}} \\
    \midrule
    \bos{}1 & 414 & 1,343 & 1,129 & 398 & 573 & 900\\
    \bos{}2 & 575 & 1,848 & 2,063 & 747 & 909 & 440\\
    \bos{}3 & 406 & 1,356 & 1,967 & 491 & 681 & 510\\
    \hline
    \cfs{}1 & 80 & 236 & 288 & 37 & 9 & -\\
    \cfs{}2 & 40 & 194 & 369 & 16 & 31 & -\\
    \cfs{}3 & 15 & 47 & 275 & 11 & 6 & -\\
    \hline
    \mantle{}1 & 133 & 13 & 2,280 & 240 & 56 & 1610 \\
    \mantle{}2 & 65 & 9 & 760 & 150 & 367 & 1370 \\
    \mantle{}3 & 96 & 10 & 1,830 & 360 & 195 & 2110\\
    \bottomrule
    \end{tabular}
\end{table}

\metadb{} has been deployed in production at \baidu{} for more than five years, unifying metadata management in distributed storage systems.
\autoref{table:production} shows nine representative \metadb{} deployments.
\bos{} has the largest cluster among the three services, with $\sim$500 nodes, managing 1.3 to 1.8 trillion records and hundreds of PBs of data, with peaks of 2M read and 0.7M write QPS.
\cfs{} and \mantle{} are smaller than \bos{} in volume, with tens to hundreds of nodes, managing up to hundreds of billions of records, supporting up to 2M read and 360K write QPS.
\metadb{} can also handle extra-large shards.
For example, the largest shards in the three \cfs{} clusters are 179\,GB, 38\,GB and 52\,GB, holding 901M, 142M and 194M records, respectively.
The last column shows the maximum gap in the number of queries between primary metadata updates and asynchronous index update requests being successfully sent to index shards, indicating that the latency of asynchronous updates is acceptable in the worst cases.

\begin{table}
    \footnotesize
    \centering
    \caption{Multi-shard transaction ratios by workload.}
    \label{table:multi_shard}
    \begin{tabular}{p{2cm}|r|r}
    \toprule
    \textbf{\textit{Workload}} &
    \textbf{\textit{GP-DB (\%)}  } &
    \textbf{\textit{MetaDB (\%)}  } \\
    \midrule
    \cfs{}1    & 100 & 0.79 \\
    \cfs{}2    & 100 & 2.37 \\
    \mantle{}1 & 100 & 4.37 \\
    \mantle{}2 & 100 & 3.13 \\
    \bottomrule
    \end{tabular}
\end{table}
\begin{figure}[t!]
    \centering
    \includegraphics[width=1\linewidth]{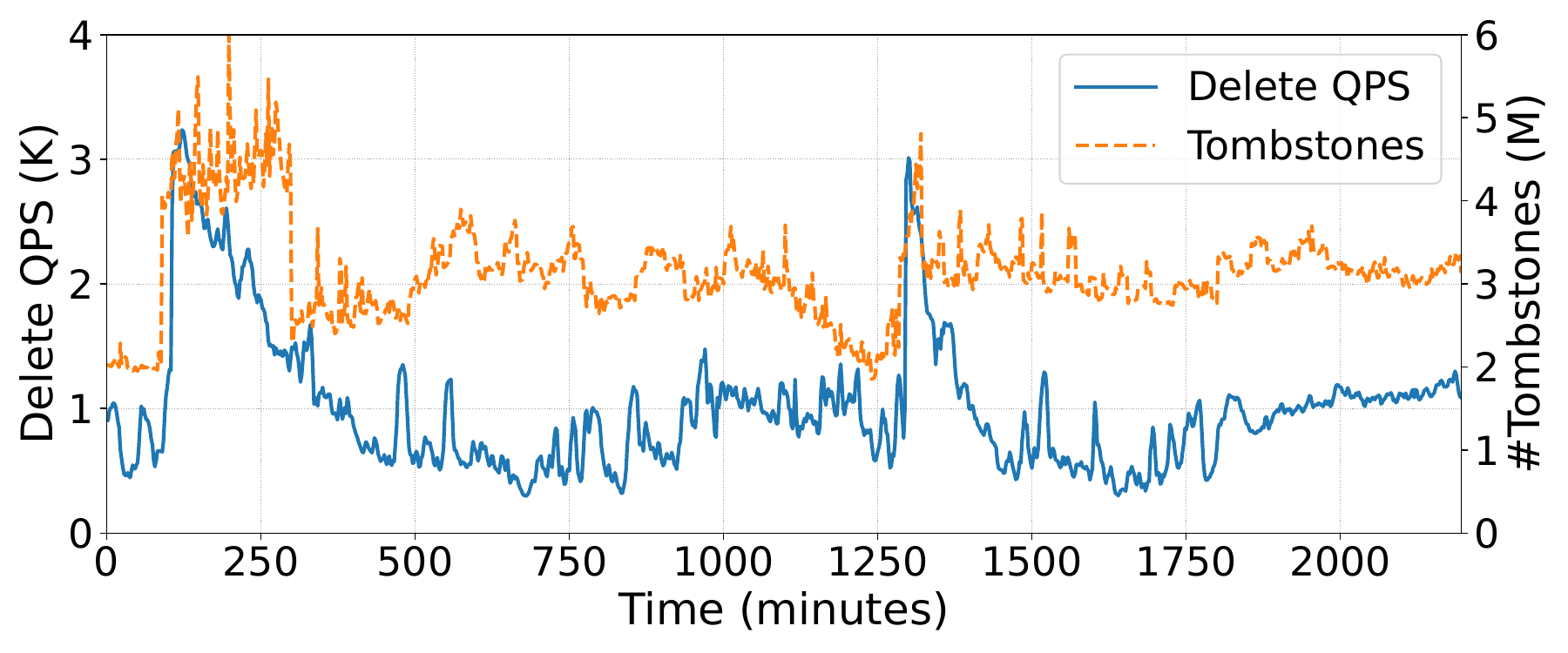}
    \caption{Tombstones and delete throughput in production.}  
    \label{fig:eval-production-deletion} 
\end{figure}

To show that \metadb{} can reduce the ratio of multi-shard transactions in production, we collect the ratio of multi-shard directory-modification transactions across four \mantle{} and \cfs{} clusters, the two services with hierarchical namespaces, and report the results in~\autoref{table:multi_shard}.
The fraction of multi-shard transactions drops from 100\% to 0.79\%--4.37\%.
For a file system built on the previous GP-DB, a single-directory modification touches records spread across multiple shards, and in Mantle it also updates a global quota index on another shard, so the modification becomes a multi-shard transaction.
In contrast, \metadb{} uses policy/size-based sharding to anchor a directory's attributes and its direct children within the same shard~(\S\ref{sec:sharding-policy}), and decouples asynchronous auxiliary updates to move quota and index updates off the transaction path~(\S\ref{sec:async-index}), so most single-directory modifications now stay within a single shard.
The remaining fraction is higher in \mantle{}, where path resolution relies on additional index entries on other shards, so \texttt{mkdir} and \texttt{rmdir} span shards as well. In \cfs{}, the only remaining multi-shard transactions come from cross-directory \rename{}, whose ratio is low ~(\autoref{fig:workload-ops}), so the fraction stays below 3\%.

\begin{figure}[t!]
    \begin{subfigure}[b]{1\linewidth}
        \includegraphics[width=\linewidth]{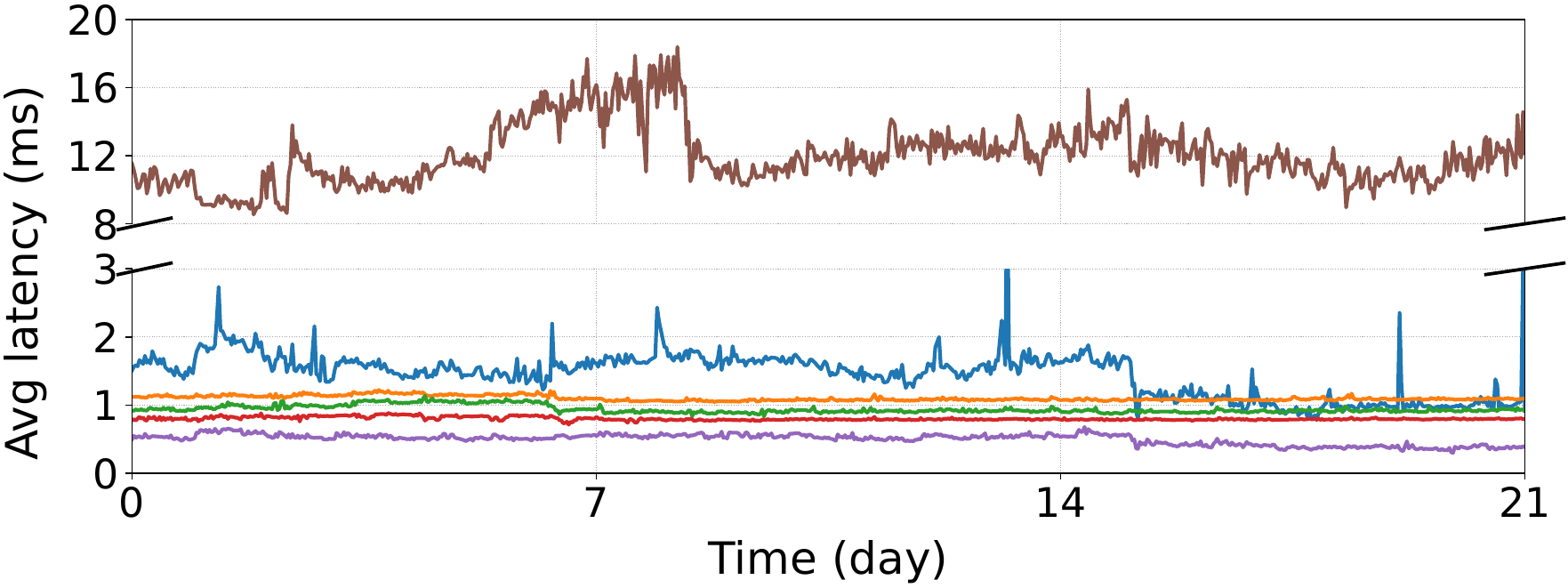}
        \caption{GP-DB}     
        \label{fig:eval4-lat-origin}
    \end{subfigure}
    \hfill
    \makebox[\linewidth][r]{%
        \begin{subfigure}[b]{0.99\linewidth}
            \includegraphics[width=\linewidth]{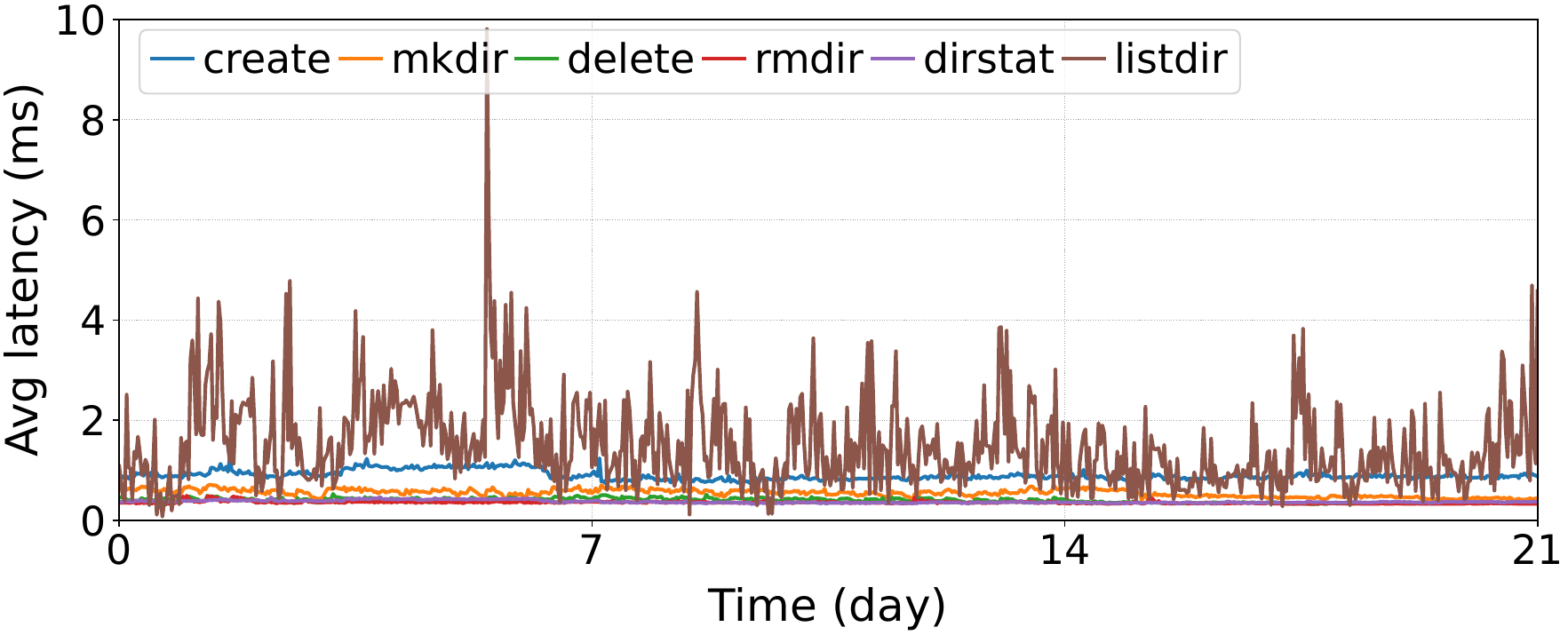}
            \caption{\metadb{}}
            \label{sec:eval4-lat-tafdb}
        \end{subfigure}%
    }
    \caption{Latency of the general-purpose database~(GP-DB) and \metadb{} in our production clusters over 21 days.}
    \label{fig:eval4-lat-trace}
\end{figure}

Next, we show that the number of tombstones is well controlled in production. \autoref{fig:eval-production-deletion} shows tombstone count and deletion throughput on a BE node that exhibits high deletion throughput in a \bos{} cluster which mainly serves ML training and search engines.
The number of tombstones remains low, peaking at $\sim$6M and varying around 3M during continuous deletions mainly between 0.5K and 3K\,QPS.
This shows that in-memory MVCC and locality-aware compaction can effectively control tombstone accumulation.

To show the effect of \metadb{} on latency in production, we trace the average latency of metadata operations on two \cfs{} clusters for 21 days and show the results in~\autoref{fig:eval4-lat-trace}.
Since most clusters are now running \metadb{}, GP-DB only runs at a small scale, with 4 BE nodes, managing 1 billion directories and files. 
The \metadb{} cluster has 50 BE nodes managing 50 billion directories and files.
Both clusters primarily run large-scale data analysis workloads, such as bioinformatics and autonomous driving.
\metadb{} shows the largest improvement for \scan{} where the typical latency decreases from around 8--18\,ms to 0.5--4\,ms.
The latencies of \create{}, \mkdir{}, \delete{}, and \rmdir{} drop from roughly 1.5\,ms, 1.1\,ms, 0.9\,ms, and 0.8\,ms under GP-DB to around 0.9\,ms, 0.5\,ms, 0.4\,ms and 0.3\,ms under \metadb{}, respectively.
\metadb{} also shows a latency decrease in \dirstat{}, where the latency decreases from around 0.5\,ms to 0.4\,ms.
These results show the effectiveness of \metadb{} in reducing operation latency in real production environments, even at a larger scale than GP-DB.
It is worth noting that these production observations are not intended as strict performance comparisons.

\section{Experiences and Lessons}
\label{sec:discussion}

\noindent\textbf{\emph{Decoupling vs. Controlled Coupling?}}
Decoupling directory management from the underlying database can increase the consistency-maintenance complexity and lead to redundant engineering.
In an early hierarchical namespace design for object storage, we kept the directory indexing subsystem separate from the database, strictly enforcing the boundary between the directory semantic layer and the database layer. 
This separation caused two problems. 
First, operations such as \texttt{mkdir} had to atomically update both the directory hierarchy in the indexing subsystem and the metadata records in the database, forcing a cross-system two-phase commit protocol with high engineering complexity. 
Second, the indexing subsystem had to implement its own replication, failure detection, and recovery, even though the database already provided these capabilities. 
We later integrated the directory indexing subsystem into \metadb{} through internal tables and native transactions, allowing directory operations and metadata updates to complete within a single ACID transaction. This removed cross-system consistency issues, reused the database’s replication and recovery mechanisms, and saved roughly six months of work for a ten-engineer team.

\noindent\textbf{\emph{Building from Scratch vs. Modifying General-Purpose Distributed Databases?}}
For metadata workloads, it is often better to build a simpler and more controllable specialized system than to deeply modify a large general-purpose database.
Generation 2~(\S\ref{sec:prev-tafdb}) showed these costs on a database we built ourselves, and AFS, our in-house HDFS-like file system, ran into the same mismatches~(\S\ref{subsec:motivation}) on CockroachDB~\cite{sigmod20-cockroachdb}, which suggests the problem is not specific to one database.
Adding metadata-specific mechanisms to such a system is costly.
Mainstream open-source distributed transactional databases typically contain millions of lines of code~\cite{vldb20-tidb, tidb, cockroachdb-github}, making invasive changes to their sharding, transaction, or storage layers expensive. 
Rather than keep modifying Generation 2, we therefore built \metadb{} from scratch with RocksDB as the local storage engine. 
The codebase is $\sim$150K lines, less than one-tenth the size of comparable open-source distributed transactional databases.
The trade-off is that when new requirements arise, we may need to implement features that a general-purpose database already provides. Because our metadata requirements and access patterns have changed little, this extra development cost has remained manageable.

\noindent
\textbf{\emph{SQL Interface vs.~Dedicated API?}}
A metadata database can keep SQL for compatibility without running a general SQL engine.
The storage services had accessed earlier metadata databases through SQL~(\S\ref{sec:prev-tafdb}), so keeping SQL eased their migration to \metadb{}.
In GP-DB, however, every request went through full parsing, semantic analysis, query optimization, and plan generation. The CPU cost was disproportionate to the metadata operations, since they use only $\sim$20 fixed SQL patterns.
\metadb{} therefore replaced the general SQL engine with a lightweight direct execution framework that maps each pattern to a storage execution path.

\noindent 
\textbf{\emph{Single Manager vs. Two-Tier Management?}}
A control plane should be partitioned before shard growth makes it a bottleneck.
Each shard has its own Raft group~\cite{multiraft}, as in CockroachDB~\cite{sigmod20-cockroachdb}, so the Manager's heartbeat, reporting, replica scheduling, and routing workload grows with the shard count.
A single cluster already holds millions of shards, and this number continues to grow.
We therefore partitioned the physical cluster into independently managed Cells, each with a Local Manager for shard management.
The Global Manager maintains a small set of mappings from contiguous key ranges to Cells for global routing.
For cross-Cell migration, it selects a range in the source Cell and coordinates with both Local Managers to move the shards in that range to the destination Cell's BE nodes.
Per-shard heartbeats and scheduling remain within each Cell, keeping the Global Manager off the per-shard control path.

\noindent 
\textbf{\emph{Central Services: Simplicity vs. Availability?}}
A central service is simple to build, but on the critical path it can affect the whole cluster when it fails or is overloaded.
In one incident, several Proxies crashed and restarted with empty routing caches, and their routing lookups overloaded the Manager.
We therefore placed MetaCache, a scalable group of cache servers, in front of the Manager to serve these lookups.
The TimeStamp Oracle~(TSO)~\cite{osdi10-percolator} in early versions of \metadb{} had a similar risk.
A TSO failure would block writes.
We replaced it with a Hybrid Logical Clock~(HLC)~\cite{opodis14-hlc}, which lets each node generate timestamps locally.

\section{Related Work}

\noindent
\textbf{Distributed Metadata Management.}
Some file systems build their own distributed metadata services.
CephFS~\cite{osdi06-ceph} and Lustre DNE~\cite{lustre-dne} assign metadata to servers at subtree granularity and split a single directory across multiple metadata servers.
CephFS also migrates subtrees according to runtime load.
IndexFS~\cite{sc14-indexfs} assigns metadata servers per directory and splits entries across servers with GIGA+~\cite{fast11-giga-plus} as directories grow.
LocoFS~\cite{sc17-locofs} separates directory metadata from file metadata to raise the throughput of file metadata operations.
InfiniFS~\cite{fast22-infinifs} colocates the parent directory's content metadata with the metadata of its child files where possible, which reduces cross-node coordination for common updates.

HopsFS~\cite{fast17-hopfs} and Tectonic~\cite{fast21-tectonic} build metadata services on general-purpose distributed databases, NDB~\cite{ndb} and ZippyDB~\cite{zippydb} respectively.
HopsFS typically partitions by parent directory ID, so the direct children of a directory reside in the same database shard.
It uses row locks for isolation and two-phase commit for atomic cross-shard updates.
When many operations update the same directory, lock contention on the parent directory's record limits throughput~\cite{fast17-hopfs, eurosys23-cfs}.
Tectonic relies on shard-local atomic operations and does not guarantee atomic cross-directory rename.
CFS~\cite{eurosys23-cfs} and Mantle~\cite{sosp25-mantle}, in contrast, run on \metadb{}, a metadata-aware database (called TafDB in both papers), which lets common single-directory operations avoid cross-shard commits and reduces contention on parent directory attributes.
The CFS and Mantle papers focus on the namespace layer.
This paper presents the core design of the database beneath them and reports cross-service production evaluation and deployment experience.

\noindent 
\textbf{Distributed Databases.}
Most existing distributed databases are general-purpose systems rather than databases specialized for metadata workloads, such as Spanner~\cite{osdi12-spanner}, CockroachDB~\cite{sigmod20-cockroachdb, cockroachdb}, FoundationDB~\cite{sigmod21-foundationDB, foundationdb}, YugabyteDB~\cite{yugabytedb}, TiDB~\cite{vldb20-tidb, tidb}, DynamoDB~\cite{atc22-dynamodb, atc23-dynampdb-transaction}, and OceanBase~\cite{vldb22-oceanbase, oceanbase}. Unlike these systems, \metadb{} is specialized around the semantics of metadata operations, including its sharding, transaction processing, and storage-engine behaviors.

\noindent
\textbf{Asynchronous Auxiliary Metadata Maintenance.}
Prior work has used asynchronous index updates in databases to improve ingestion throughput, reduce synchronization overhead, or improve overall scalability~\cite{icnidc12-ais, vldb20-hydralist, osdi23-scaledb, hudi-async-index}.
These works primarily focus on indexing performance or database scalability. In contrast, \metadb{} uses asynchronous maintenance to restructure the metadata transaction path. 
It removes auxiliary metadata updates from the primary transaction by exploiting their relaxed timeliness requirements, thereby avoiding unnecessary multi-shard coordination. 

\noindent \textbf{LSM-Tree-Based KV Store Deletion.} Marked deletion degrades LSM-tree performance~\cite{tos21-rocksdb, rocksdb}. 
Existing works enforce latency-bounded deletion for performance~\cite{sigmod20-lethe} or privacy~\cite{sigmod23-acheron,tods23-deletion-lsm}. 
However, these \textit{retention-based} strategies require intrusive engine modifications and incur high write amplification by eagerly compacting cold data. 
In contrast, \metadb{} addresses this from the upper layer via \textit{locality-aware compaction}. 
Instead of rigid time windows, it targets deletion hotspots based on density. 
This triggers compaction only when necessary for read performance, avoiding the I/O overhead of time-driven approaches.

\section{Conclusion}
\label{sec:conclusion}

This paper presents \metadb{}, a metadata-aware distributed database. 
\metadb{} is co-designed around the file-system structural, semantic, and dynamic properties of metadata workloads. 
Our evaluation and deployment experience at \baidu{} show that this design substantially improves both performance and efficiency. 
\metadb{} achieves 2.2$\times$--4.1$\times$ the throughput of the previous general-purpose database, halves the number of metadata machines in production and sustains stable low latency for range scans under continuous bulk deletions. 
These results suggest that, for metadata management in storage systems, exposing and exploiting workload-specific structure inside the database is a key design principle for achieving both high performance and efficiency.

\section*{Acknowledgments}
We thank the anonymous reviewers and Jiancong Zheng for their insightful feedback.
Zebin Ren is funded by the Dutch Research Council under grant numbers OCENW.KLEIN.561 and OCENW.KLEIN.209.
We thank Meng Wang, Yongqiang Yang, Miao Yu, Xianhui Niu, Linjiang Lian, Guangjun Xie, Baolei Jia, and Pengfei Zheng for their contributions to the development of \metadb{}. 
We also thank Rurue Ma, Chun Zhao, Yuhang Mu, and Liping Liu for their contributions to the development of \metadb{}'s predecessor system.

\bibliographystyle{plain}
\bibliography{main}

\clearpage

\appendix
\section{Performance Comparison with \tidb{}}
\label{a1:eval-gp-db}

The \tidb{} version evaluated here requires a newer operating-system version than the one deployed in our production cluster.
Thus, we set up a separate environment with a different set of machines for the experiments in this appendix.
The \metadb{} cluster is configured with 4 BE nodes, 3 Manager nodes, and 3 Proxy nodes, and the~\tidb{} cluster is configured with 4 TiKV nodes, 3 PD nodes, and 3 \tidb{} nodes for \tidb{} v8.1.
Both clusters contain the same type of machines, where each node is configured with two Intel Xeon 4314 processors~(32 cores), 128\,GB of DDR4-3200 memory, 8$\times$4\,TB NVMe SSDs, and two 25 Gbps network connections.
Each node runs CentOS 7.5 (Kernel 4.14.0).

\begin{figure} 
\centering

\begin{minipage}{0.22\textwidth}
    \centering
    \includegraphics[width=\linewidth]{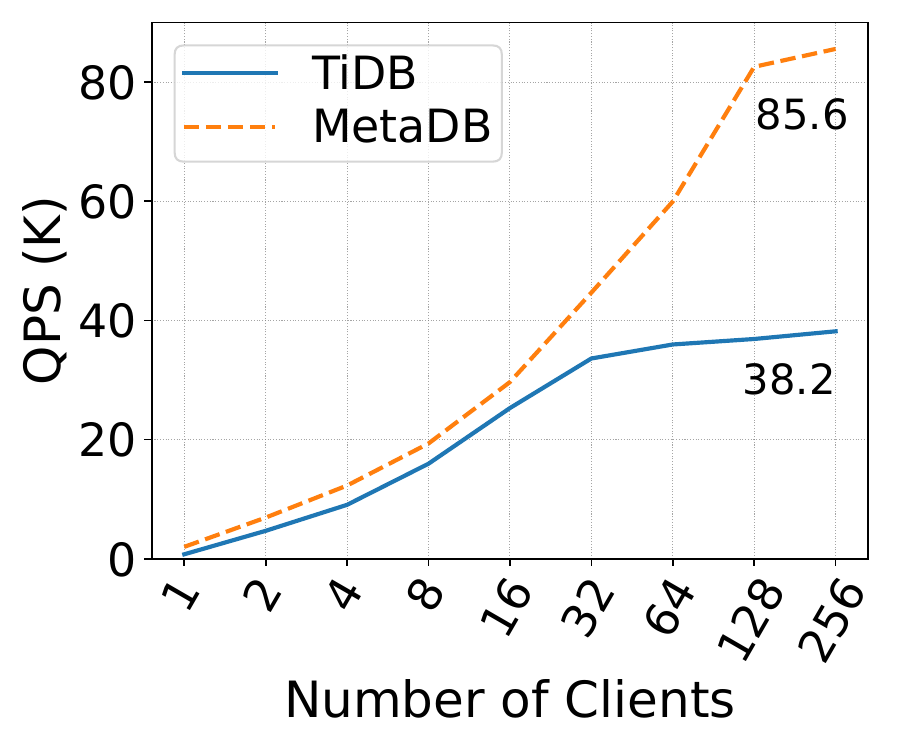}
    \caption{Operation throughput with index updates with different concurrencies.}
    \label{fig:eval-tidb-single-multi}
\end{minipage}
\hfill
\begin{minipage}{0.22\textwidth}
    \centering
    \includegraphics[width=\linewidth]{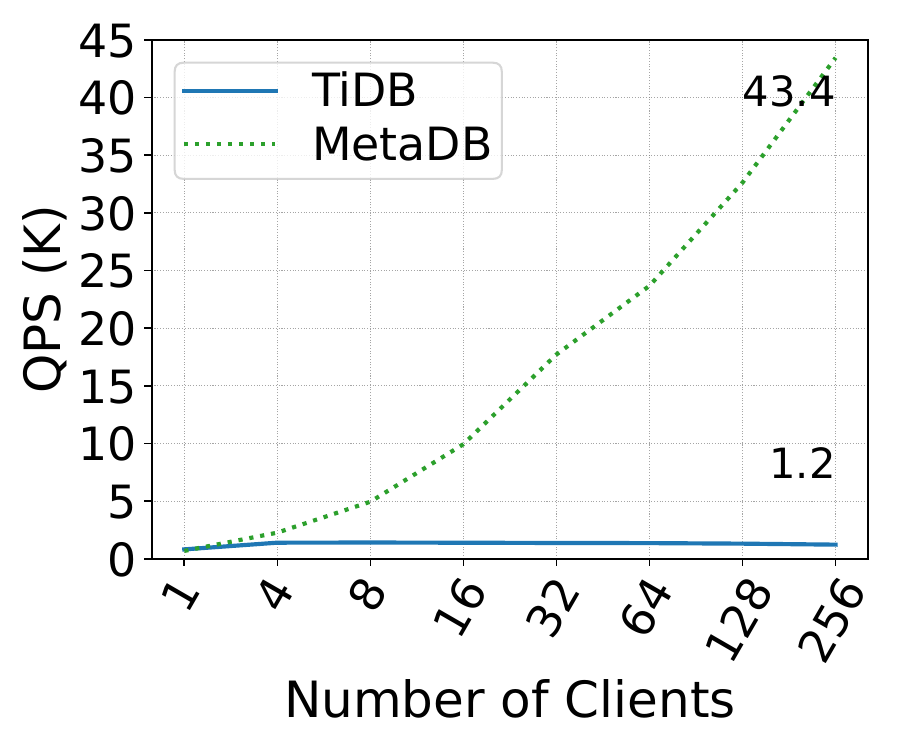}
    \caption{Throughput of operations under the same directory with different concurrencies.}
    \label{fig:eval-tidb-conflict}
\end{minipage}

\end{figure}

\begin{figure}
    \centering
    \includegraphics[width=0.9\linewidth]{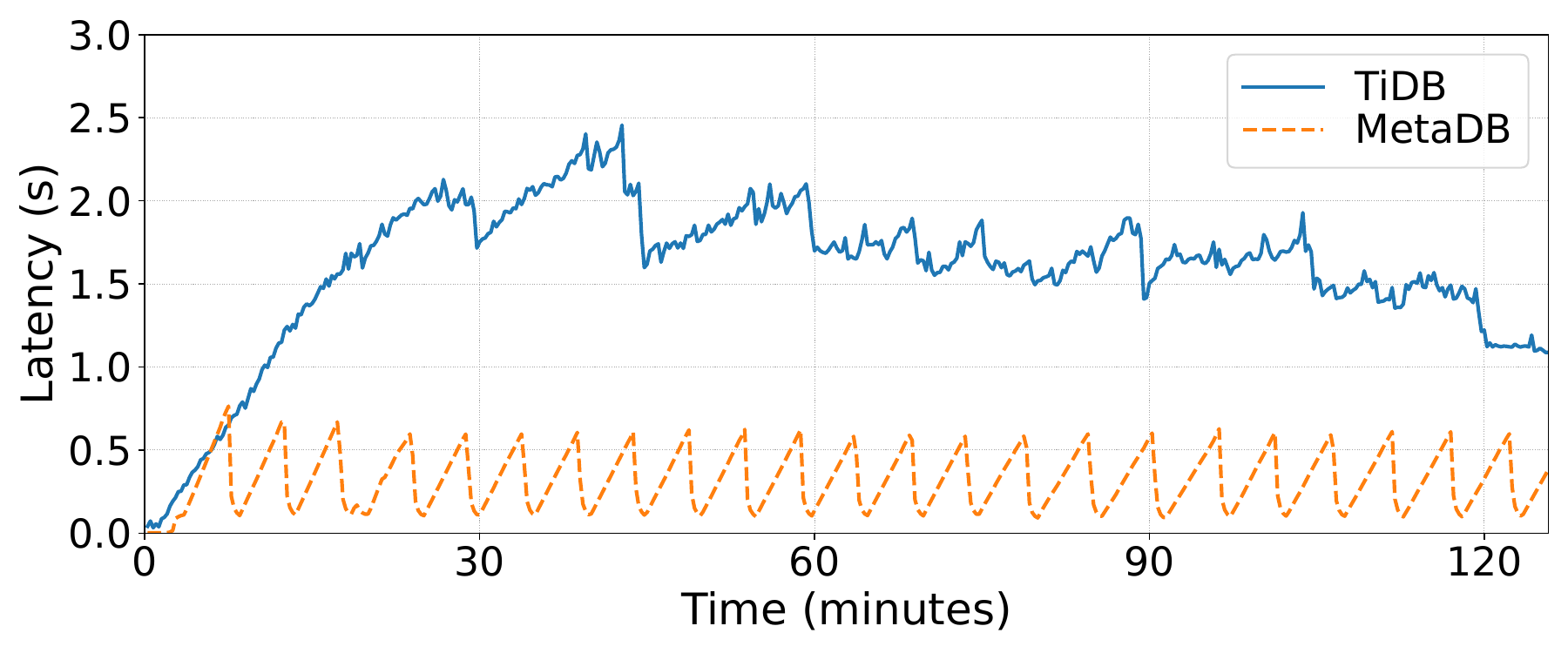}
    \caption{\scan{} latency with concurrent consecutive bulk deletion between TiDB and \metadb{}.}      
    \label{fig:eval-tidb-deletion}  
\end{figure}

\autoref{fig:eval-tidb-single-multi} compares the throughput of \create{} with index updates between \tidb{} and \metadb{}, where \tidb{} updates the index atomically with the primary metadata update within the same transaction and \metadb{} updates the index asynchronously, and each client sends four concurrent requests.
The throughput of \metadb{} with 256 clients is 85.6K\,QPS, which is 2.2$\times$ \tidb{}'s throughput with 256 clients.
This shows that our asynchronous index framework can deliver higher throughput than updating indexes atomically using transactions.

Next, we compare the throughput between \tidb{} and \metadb{} with concurrent \mkdir{} under the same directory where each client sends four concurrent requests and present their throughput in~\autoref{fig:eval-tidb-conflict}.
\metadb{} has better scalability than \tidb{}, achieving 36.2$\times$ the peak throughput of \tidb{} for operations that modify the same row~(i.e., the parent directory's attribute).
This demonstrates that \metadb{} can achieve higher throughput than general-purpose databases under contention from concurrent operations within the same
directory.

\autoref{fig:eval-tidb-deletion} compares the latency of \scan{} running concurrently with object deletions at 5K QPS on \metadb{} and \tidb{}. \metadb{}’s \scan{} latency stays below 0.7\,s, which is lower and more stable than \tidb{}’s latency that peaks at 2.5\,s and stays between 1\,s and 2.5\,s after the tombstones have accumulated for 30 minutes.
This demonstrates that \metadb{} is more effective at controlling \scan{} latency during consecutive bulk deletions than \tidb{}.

\end{document}